\documentclass{aa}
\usepackage{graphicx}

\def\mb#1{\setbox0=\hbox{$#1$}\kern-.025em\copy0\kern-\wd0
\kern-0.05em\copy0\kern-\wd0\kern-.025em\raise.0233em\box0}

\begin{document}

   \title{Dynamical stability of collisionless stellar systems \\
   and barotropic stars: the nonlinear Antonov first law}

\titlerunning{Dynamical stability of stars and galaxies}

   \author{P.H. Chavanis}

\institute{ Laboratoire de Physique Th\'eorique (UMR 5152 du CNRS), Universit\'e Paul
Sabatier, 118 route de Narbonne 31062 Toulouse, France\\
\email{chavanis@irsamc.ups-tlse.fr}}

   \date{\today}

\abstract{We complete previous investigations on the dynamical
stability of barotropic stars and collisionless stellar systems. A
barotropic star that minimizes the energy functional at fixed mass is
a nonlinearly dynamically stable stationary solution of the
Euler-Poisson system. Formally, this minimization problem is similar
to a condition of ``canonical stability'' in thermodynamics. A
stellar system that maximizes an $H$-function at fixed mass and
energy is a nonlinearly dynamically stable stationary solution of the
Vlasov-Poisson system. Formally, this maximization problem is similar
to a condition of ``microcanonical stability'' in
thermodynamics. Using a {\it thermodynamical analogy}, we provide a
derivation and an interpretation of the nonlinear Antonov first law in
terms of ``ensembles inequivalence'': a spherical stellar system with
$f=f(\epsilon)$ and $f'(\epsilon)<0$ is nonlinearly dynamically stable
with respect to the Vlasov-Poisson system if the corresponding
barotropic star with the same equilibrium density distribution is
nonlinearly dynamically stable with respect to the Euler-Poisson
system. This is similar to the fact that ``canonical stability
implies microcanonical stability'' in thermodynamics. The converse is
wrong in case of ``ensembles inequivalence'' which is generic for
systems with long-range interactions like gravity. We show that
criteria of nonlinear dynamical stability can be obtained very simply
from purely graphical constructions by using the method of series of
equilibria and the turning point argument of Poincar\'e, as in
thermodynamics.

  }

\maketitle

\section{Introduction}
\label{sec_introduction}

The {dynamical stability} of stars and galaxies is an important topic
in astrophysics. A large number of results have been obtained in the
past and are collected in Binney \& Tremaine (1987). Most
of these results are based on a linearization of the Euler-Poisson and
Vlasov-Poisson systems around a stationary solution and on the study
of the evolution of the perturbation which is decomposed in normal
modes $e^{\sigma t}h_{\sigma}({\bf x})$.  In practice, one has to work
out an eigenvalue equation determining the complex pulsation $\sigma$
and the corresponding eigenfunction $h_{\sigma}({\bf x})$.  This
approach provides a condition of {\it linear dynamical stability} when
$R_e(\sigma)\le 0$ for all $\sigma$ and a condition of instability
otherwise. It has been found that a spherical stellar system with DF
$f=f(\epsilon)$ and $f'(\epsilon)<0$, where
$\epsilon=v^{2}/2+\Phi({\bf r})$ is the individual energy, is linearly
dynamically stable with respect to the Vlasov-Poisson system if the
barotropic gas with the same equilibrium density distribution is
linearly dynamically stable with respect to the Euler-Poisson
system. This forms the Antonov first law (1960) and explains why the
stability of stellar systems and gaseous systems are often treated in
parallel although these systems are physically different (Binney \&
Tremaine 1987).

In this paper, we shall tackle these dynamical stability problems by a
different approach, using optimization principles.  These optimization
principles provide conditions of {\it nonlinear dynamical stability}
for barotropic stars and spherical galaxies. This is different from
the {\it thermodynamical stability} of stellar systems (Padmanabhan
1990). In the thermodynamical context, we maximize the Boltzmann
entropy $S_{B}[f]=-\int {f\over m}\ln {f\over m}d{\bf r}d{\bf v}$ at
fixed mass $M[f]$ and energy $E[f]$ in order to obtain the {\it most
probable} equilibrium state ($f\sim e^{-\beta m\epsilon}$) of a $N$-body
Hamiltonian system resulting from a ``collisional'' relaxation for
$t\gg t_{relax}$, where $t_{relax}\sim (N/\ln N)t_{D}$ is the
Chandrasekhar relaxation time ($t_{D}$ is the dynamical time). In that
case, the Boltzmann entropy is obtained from a combinatorial analysis
(Ogorodnikov 1965). On the other hand, the maximization of {\it any}
$H$-function $S[f]=-\int C(f)d{\bf r}d{\bf v}$, where $C$ is convex
(i.e. $C''>0$), at fixed mass and energy determines a {\it nonlinearly
dynamically stable} stationary solution $f=f(\epsilon)$ with
$f'(\epsilon)<0$ of the Vlasov-Poisson system which describes
collisionless stellar systems for $t\ll t_{relax}$. Formally, these
maximization problems are mathematically similar (the second one
involving a larger class of functionals than the Boltzmann functional)
but their physical content and interpretation are completely
different: the first one refers to thermodynamics (collisional regime)
while the second refers to dynamics (collisionless regime). However,
due to their mathematical similarity, we can develop a {\it
thermodynamical analogy} (Sec. \ref{sec_thanal}) to analyze the
nonlinear dynamical stability problem. In this analogy, the criterion
of nonlinear dynamical stability for spherical stellar systems with
$f=f(\epsilon)$ and $f'(\epsilon)<0$ described by the Vlasov-Poisson
system (Sec. \ref{sec_ndss}) is similar to a criterion of
thermodynamical stability in the microcanonical ensemble. We have also
found that the criterion of nonlinear dynamical stability for
barotropic stars described by the Euler-Poisson system
(Sec. \ref{sec_ndsg}) is similar to a criterion of thermodynamical
stability in the canonical ensemble.  Using this thermodynamical
analogy, we provide a derivation and an interpretation of the
nonlinear Antonov first law (Sec. \ref{sec_anto}) in terms of
``ensembles inequivalence'': a spherical stellar system with
$f=f(\epsilon)$ and $f'(\epsilon)<0$ is nonlinearly dynamically stable
with respect to the Vlasov-Poisson system if the corresponding
barotropic star with the same equilibrium density distribution is
nonlinearly dynamically stable with respect to the Euler-Poisson
system.  This is similar to the fact that ``canonical stability
implies microcanonical stability'' in thermodynamics (see, e.g., Ellis
et al. 2000, Bouchet \& Barr\'e 2005). The converse is wrong in case
of ``ensembles inequivalence'' which is generic for systems with
long-range interactions like gravity. We show that criteria of
nonlinear dynamical stability can be obtained very simply from purely
graphical constructions by using the method of series of equilibria
and the turning point argument of Poincar\'e (Secs. \ref{sec_poly} and
\ref{sec_otherdim}), like in thermodynamics
(Sec. \ref{sec_thermo}). This will open a connection between studies
of dynamical and thermodynamical stability
(Sec. \ref{sec_main}). These results were sketched at the end of a
preceding paper (Chavanis 2003a) but we think that it is important to
re-discuss them here in a more logical way with amplification.

Optimization principles for self-gravitating systems have also been
developed in connection with a notion of ``generalized
thermodynamics'' (Tsallis 1988). It has been observed in many systems
with long-range interactions (self-gravitating systems,
two-dimensional turbulence, HMF model,...) that coherent structures
(galaxies, jets and vortices, clusters,...) form spontaneously and
persist for extremely long times (see Chavanis 2002c). Since
these meta-equilibrium states, or quasi-stationary states (Q.S.S.),
are not described by the Boltzmann distribution in general, some
authors have suggested that they might be described by a generalized
thermodynamics. In particular, it has been proposed to replace the
Boltzmann entropy $S_{B}[f]$ by the Tsallis entropy $S_{q}[f]=-{1\over
q-1}\int (f^{q}-f)d{\bf r}d{\bf v}$. In that respect, the coherent
structures that quickly form in long-range systems (in the
collisionless regime) are claimed to maximize the Tsallis entropy at
fixed mass and energy (Boghosian 1996, Latora et al. 2002). In the
case of self-gravitating systems, this generalized thermodynamical
interpretation has been defended by Taruya \& Sakagami (2003). We
think, however, that this approach stems from a mis-interpretation
(Chavanis 2003a,2005b). Metaequilibrium states such as galaxies
result from an incomplete violent relaxation (Lynden-Bell 1967) and
are stable stationary solutions of the Vlasov equation on the
coarse-grained scale. The Tsallis functional
$H_{q}[\overline{f}]=-{1\over q-1}\int
(\overline{f}^{q}-\overline{f})d{\bf r}d{\bf v}$, expressed as a
function of the coarse-grained DF $\overline{f}({\bf r},{\bf v})$,
is a particular $H$-function in the sense of Tremaine et al.  (1986),
not a thermodynamical entropy. Its extremization at fixed mass and
energy leads to stellar polytropes (Sec. \ref{sec_poly}) which are
particular stationary solutions of the Vlasov equation that form
simple mathematical models of spherical stellar systems. Furthermore,
its maximization provides a condition of nonlinear dynamical stability
for the Vlasov-Poisson system, not a condition of ``generalized
thermodynamical stability''. The resemblance of the criterion of
nonlinear dynamical stability with a criterion of ``generalized
thermodynamical stability'' is essentially the mark of a {\it
thermodynamical analogy} (Chavanis 2003a).  In fact, the previous
authors did not consider the Vlasov equation (which is capital in our
interpretation) and tried to build a generalized thermodynamics in
order to deal with observed non-Boltzmannian quasi-stationary
states (Q.S.S.). Now, considering the Vlasov equation, there is no
fundamental reason why the Q.S.S. resulting from incomplete violent
relaxation should be described by stellar polytropes (Tsallis
distribution). These form just a {\it particular} class of steady
solutions of the Vlasov equation. In practice, it is extremely
difficult to {\it predict} the metaequilibrium state reached by the
system in case of incomplete relaxation and we are led to construct
dynamically stable stationary solutions of the Vlasov equation so as
to {\it reproduce} observations a posteriori (Binney \& Tremaine
1987).  The Vlasov equation admits an infinite number of stationary
solutions and only stable solutions must be considered. For spherical
systems with $f=f(\epsilon)$ and $f'(\epsilon)<0$, a powerful
criterion of nonlinear dynamical stability is given by the above
maximization principle. A connection with theories of incomplete
violent relaxation that try to predict the DF reached by the system
(Lynden-Bell 1967, Stiavelli \& Bertin 1987, Chavanis et al. 1996)
is discussed in Sec. \ref{sec_phen}.

\section{Dynamical stability of stars and galaxies}
\label{sec_gas}

\subsection{Nonlinear dynamical stability of barotropic stars}
\label{sec_ndsg}

We consider a self-gravitating gaseous medium (star) described by the Euler-Poisson
system
\begin{equation}
{\partial\rho\over\partial t}+\nabla \cdot (\rho {\bf u})=0,
\label{ep1}
\end{equation}
\begin{equation}
{\partial {\bf u}\over\partial t}+({\bf u}\cdot \nabla) {\bf u}=-{1\over\rho}\nabla p-\nabla\Phi,
\label{ep2}
\end{equation}
\begin{equation}
\Delta\Phi=4\pi G\rho.
\label{ep3}
\end{equation}
We shall restrict ourselves to the case of a {\it barotropic} gas for
which the equation of state $p=p(\rho)$ depends only on the
density. With this assumption, the Euler-Poisson system is closed. The
total energy of the fluid is
\begin{eqnarray}
{\cal W}[\rho,{\bf
u}]=\int\rho\int^{\rho}{p(\rho')\over\rho^{'2}}d\rho'd{\bf
r} +{1\over 2}\int \rho\Phi d{\bf r}+\int \rho
{{\bf u}^{2}\over 2}d{\bf r}.\nonumber\\
 \label{ep4}
\end{eqnarray}
The first term is the work $-p(\rho)d(1/\rho)$ done in compressing
it from infinite dilution, the second term is the gravitational
energy and the third term is the kinetic energy associated with
the mean motion.

It is straightforward to verify that the energy functional (\ref{ep4})
is conserved by the Euler-Poisson system ($\dot {\cal
W}=0$). Therefore, a minimum of ${\cal W}$ at fixed mass determines a
stationary solution of the Euler-Poisson system which is (formally)
nonlinearly dynamically stable in the sense of Holm {et al. } 
(1985). Physically, this means that a small perturbation from the
minimum remains close (in some norm) to the minimum.  We are led
therefore to consider the minimization problem
\begin{eqnarray}
{\rm Min}\ \lbrace {\cal W}[\rho,{\bf u}]\quad |\ M[\rho]=M\rbrace
. \label{ej4}
\end{eqnarray}
Cancelling the first order variations of Eq. (\ref{ep4}) at fixed mass, we
obtain ${\bf u}={\bf 0}$ and the condition of hydrostatic
equilibrium
\begin{equation}
\nabla p=-\rho\nabla\Phi.
\label{ep5}
\end{equation}
Therefore, extrema of ${\cal W}$ correspond to stationary solutions of
the Euler-Poisson system (\ref{ep1})-(\ref{ep3}). Combining the
condition of hydrostatic equilibrium (\ref{ep5}) and the equation of
state $p=p(\rho)$, we get
\begin{eqnarray}
\int^{\rho}{p'(\rho')\over\rho'}d\rho'=-\Phi,
\label{ep6}
\end{eqnarray}
so that $\rho$ is a function of $\Phi$ that we note
$\rho=\rho(\Phi)$. Considering now the second order variations, the
condition of nonlinear dynamical stability is
\begin{eqnarray}
\delta^{2}{\cal W}={1\over 2}\int \delta\rho\delta\Phi d{\bf r}+\int {p'(\rho)\over 2\rho}(\delta\rho)^{2}d{\bf r}\ge 0,
\label{ep7}
\end{eqnarray}
for all perturbations that conserve mass, i.e. $\int \delta\rho \
d{\bf r}=0$. We note that the second integral in Eq.  (\ref{ep7})
can be written in a more conventional form by using
$p'(\rho)/\rho=-1/\rho'(\Phi)$ resulting from Eq. (\ref{ep6}).

Let us consider explicitly some particular examples. For an isothermal
gas $p=\rho {k_{B}T\over m}$, we find that
\begin{eqnarray}
\rho=A' e^{-{m\Phi\over k_{B}T}},
 \label{ep7a}
\end{eqnarray}
\begin{eqnarray}
{\cal
W}=k_{B}T\int {\rho\over m}\ln {\rho\over m}\, d{\bf r}
+{1\over 2}\int \rho\Phi\, d{\bf r}+\int \rho {{\bf u}^{2}\over
2}\,d{\bf r}.
\label{ip1}
\end{eqnarray}
On the other hand, for a polytropic gas $p=K\rho^{\gamma}$, we have
\begin{eqnarray}
\rho=\biggl\lbrack \lambda-{\gamma-1\over K\gamma}\Phi\biggr
\rbrack^{1\over \gamma-1},
\label{ip5}
\end{eqnarray}
\begin{equation}
{\cal
W}={K\over \gamma-1}\int (\rho^{\gamma}-\rho)\,d{\bf r}
+{1\over 2}\int \rho\Phi \,d{\bf r}+\int \rho {{\bf u}^{2}\over
2}\,d{\bf r}.
\label{ip4}
\end{equation}
For $\gamma\rightarrow 1$, we recover the expressions (\ref{ep7a}) and
(\ref{ip1}) of the isothermal gas.  The minimization problem
(\ref{ej4}) has been studied in Chavanis (2002a,b,2003a) for an
isothermal and a polytropic equation of state (the system can be
self-confined or confined by a box or by an external pressure).  It is
shown that the condition of nonlinear dynamical stability coincides
with the condition of linear dynamical stability (see also Appendix
\ref{sec_marg}). On the other hand, for isothermal systems, the energy
functional ${\cal W}[\rho,{\bf 0}]$ coincides with the Boltzmann free
energy $F_{B}[\rho]=E[\rho]-TS_{B}[\rho]$ (see Chavanis 2002a for
details). Therefore, the condition of nonlinear dynamical stability
with respect to the Euler-Poisson system is the same as the condition
of thermodynamical stability in the canonical ensemble.

\subsection{Nonlinear dynamical stability of stellar systems}
\label{sec_ndss}

We consider a stellar system (galaxy) described by the Vlasov-Poisson system
\begin{equation}
{\partial f\over\partial t}+{\bf v}\cdot {\partial f\over\partial {\bf r}}+{\bf F}\cdot {\partial f\over\partial {\bf v}}=0,\label{vh0}
\end{equation}
\begin{equation}
\Delta\Phi=4\pi G\int f d{\bf v},\label{vh1}
\end{equation}
where ${\bf F}=-\nabla\Phi$ is the self-consistent gravitational field
produced by the stars. The Vlasov description assumes that the
evolution of the system is encounterless. This is a very good
approximation for most stellar systems because the relaxation time is
in general larger than the age of the universe by several orders of
magnitude (Binney \& Tremaine 1987). The Vlasov equation conserves the
total mass $M=\int f d{\bf r}d{\bf v}$, the total energy
$E={1\over 2}\int f v^{2}d{\bf r}d{\bf v}+{1\over 2}\int
\rho\Phi d{\bf r}$ and an infinite class of functionals called the
Casimirs that are defined by $I_{h}=\int h(f) d{\bf r}d{\bf
v}$ for any function $h$. We shall be particularly concerned with a special class
of Casimirs of the form
\begin{equation}
S[f]=-\int C(f)d{\bf r}d{\bf v},\label{vh2}
\end{equation}
where $C$ is an arbitrary convex function, i.e. $C''>0$. 
Since $S$, $M$ and $E$ are conserved by the
collisionless dynamics, the maximization problem
\begin{equation}
\label{vh3} {\rm Max}\  \lbrace S[f]\quad |\ E[f]=E, M[f]=M\rbrace,
\end{equation}
determines a nonlinearly dynamically stable stationary solution of the
Vlasov-Poisson system. We note that $F[f]=E[f]-TS[f]$, where $T$ is an
arbitrary positive constant (we shall explain later why we restrict
$T$ to positive values), is also conserved by the Vlasov
equation. This is called a Casimir-Energy  functional. Therefore, a
minimum of $F$ at fixed mass $M$ also determines a nonlinearly
dynamically stable stationary solution of the Vlasov-Poisson
system. This criterion can be written
\begin{equation}
\label{vh4} {\rm Min}\  \lbrace F[f]\quad |\ M[f]=M\rbrace.
\end{equation}
Instead of minimizing $F$ at fixed mass, it proves more convenient to
maximize the functional $J=S-\beta E$ at fixed mass, where
$\beta=1/T$. We shall often work with the functional $J$ in the
following. Similar criteria of nonlinear dynamical stability have been
introduced by Ellis et al. (2002) in the context of two-dimensional
hydrodynamics described by the 2D Euler-Poisson system. The
optimization problem (\ref{vh4}) corresponds to the formal nonlinear
dynamical stability criterion of Holm et al. (1985) and the
optimization problem (\ref{vh3}) corresponds to the refined dynamical
stability criterion of Ellis {et al.} (2002). These criteria are in
general not equivalent for systems with long-range interactions, and
this is similar to a notion of ``ensembles inequivalence'' in
thermodynamics (see Sec. \ref{sec_thanal}). The criterion of Ellis et
al. (2002) refines the nonlinear dynamical stability theorems of
Arnold (1966) who first introduced the Casimir-Energy method in 2D
hydrodynamics. Inspired by these studies, we have applied the same
criteria (\ref{vh3}) and (\ref{vh4}) to study the nonlinear dynamical
stability of collisionless stellar systems described by the
Vlasov-Poisson system (Chavanis 2003a,b, Chavanis \& Sire
2004). Similar optimization principles have been introduced
independently by Guo \& Rein (1999,2001) in the mathematical
literature. They first considered the minimization of a Casimir-Energy
functional $E-S$ at fixed mass $M$ (Guo \& Rein 1999) and later
obtained a refined criterion of nonlinear dynamical stability by
minimizing the energy $E$ at fixed Casimir $S$ and mass $M$ (Guo \&
Rein 2001). The first criterion is equivalent to minimizing $F$ at
fixed mass (optimization problem (\ref{vh4})) and the second criterion
is equivalent to maximizing $S$ at fixed energy $E$ and mass $M$
(optimization problem (\ref{vh3})).

Using Lagrange multipliers and writing respectively
\begin{equation}
\label{var1car}
\delta S-\beta\delta E-\alpha\delta M=0,
\end{equation}
or
\begin{equation}
\label{var2} \delta J-\alpha\delta M=0,
\end{equation}
the {critical points} of the variational problems
(\ref{vh3}) and (\ref{vh4}) are both given by
\begin{equation}
\label{vh6}
C'(f)=-\beta\epsilon-\alpha,
\end{equation}
where $\epsilon={v^{2}\over 2}+\Phi({\bf r})$ is the energy of a
particle by unit of mass. Since $C'$ is a monotonically increasing
function of $f$, we can inverse this relation to obtain
\begin{equation}
\label{var1}
f=F(\beta\epsilon+\alpha)
\end{equation}
where $F(x)=(C')^{-1}(-x)$. Therefore, the {\it critical points} of
the optimization problems (\ref{vh3}) and (\ref{vh4}) are the same.
They correspond to distribution functions of the form $f=f(\epsilon)$
which depend only on the energy. According to the Jeans theorem, these
DF are particular stationary solutions of the Vlasov-Poisson system
which form a sub-class of spherical galaxies (the general DF of
spherical galaxies depends on energy $\epsilon$ and angular momentum
${\bf J}={\bf r}\times {\bf v}$ or $J=|{\bf r}\times {\bf v}|$).  From
Eq.  (\ref{vh6}), we have the identity $f'(\epsilon)=-\beta/C''(f)$
showing that $f(\epsilon)$ is a monotonic function. Since
$f(\epsilon)$ should be decreasing at least for large energies, this
imposes $\beta=1/T>0$. Then, we conclude that $f'(\epsilon)< 0$ for
all $\epsilon$. Inversely, any DF of the form $f=f(\epsilon)$ with
$f'(\epsilon)< 0$ is a critical point of a certain Casimir $S$ of the
form (\ref{vh2}) at fixed mass and energy.

Furthermore, any DF of the form $f=f(\epsilon)$ with $f'(\epsilon)<0$
which satisfies the optimization problems (\ref{vh3}) or (\ref{vh4})
is nonlinearly dynamically stable with respect to the Vlasov-Poisson
system. The condition of nonlinear dynamical stability provided by the
criterion (\ref{vh3}) is
\begin{eqnarray}
\delta^{2}J=-\int C''(f){(\delta f)^{2}\over 2}d{\bf r}d{\bf v}-{1\over 2}\beta\int \delta\rho\delta\Phi d{\bf r}\le 0,\nonumber\\
\forall \delta f \quad |\quad  \delta E=\delta M=0, \qquad\qquad\qquad
\label{vh7}
\end{eqnarray}
and the condition of nonlinear dynamical stability provided by the
criterion (\ref{vh4}) is
\begin{eqnarray}
\delta^{2}J=-\int C''(f){(\delta f)^{2}\over 2}d{\bf r}d{\bf v}-{1\over 2}\beta\int \delta\rho\delta\Phi d{\bf r}\le 0,\nonumber\\
\forall \delta f \quad |\quad  \delta M=0.\qquad\qquad\qquad
\label{vh8}
\end{eqnarray}
The first integral can be written in a more conventional form by
using $f'(\epsilon)=-\beta/C''(f)$. We note the importance of the
first order constraints in the criteria (\ref{vh7}) and (\ref{vh8}).
If condition (\ref{vh8}) is satisfied for all perturbations that
conserve mass, it is a fortiori satisfied for perturbations that
conserve mass {\it and} energy. Therefore, condition (\ref{vh8})
implies condition (\ref{vh7}), but not the opposite. This is similar
to the fact that canonical stability implies microcanonical
stability in thermodynamics (but not the opposite). We conclude that
the stability criterion (\ref{vh3}) is more refined than the
stability criterion (\ref{vh4}): if a collisionless stellar system
with $f=f(\epsilon)$ and $f'(\epsilon)<0$ satisfies (\ref{vh3}) or
(\ref{vh4}), it is nonlinearly dynamically stable; however, if it
does not satisfy (\ref{vh4}), it can be nonlinearly dynamically
stable provided that it satisfies (\ref{vh3}). This means that we
can ``miss'' stable solutions if we use just the optimization
problem (\ref{vh4}) to construct stable stationary solutions of the
Vlasov-Poisson system of the form $f=f(\epsilon)$ with
$f'(\epsilon)<0$. The problem (\ref{vh3}) is richer and allows to
construct a larger class of nonlinearly dynamically stable models of
spherical galaxies. In particular, we shall see that stellar polytropes of
index $3<n<5$ satisfy the criterion (\ref{vh3}) but not the
criterion (\ref{vh4}).

\section{Thermodynamics of self-gravitating systems and series of equilibria}
\label{sec_thermo}

\subsection{Classical particles}
\label{sec_class}

Optimization problems of the form (\ref{vh3}) and (\ref{vh4}) were first
studied in relation with the thermodynamics of stellar systems
(Antonov 1962, Lynden-Bell \& Wood 1968) where $S$ is the
Boltzmann entropy
\begin{eqnarray}
S_{B}[f]=-\int {f\over m}\ln {f\over m}d{\bf r}d{\bf v},
\label{thss1}
\end{eqnarray}
which can be obtained by a combinatorial analysis (Ogorodnikov 1965).  In this
thermodynamical context, the criterion  (\ref{vh3}) is a criterion of
{\it microcanonical stability} and the criterion (\ref{vh4}) is a criterion of
{\it canonical stability} where $F_{B}[f]=E[f]-TS_{B}[f]$ is the Boltzmann
free energy and $T$ the thermodynamical temperature. The critical
points of $S_{B}$ or $F_{B}$ (with appropriate constraints) correspond
to the mean-field Maxwell-Boltzmann distribution
\begin{eqnarray}
f=A e^{-\beta m({v^{2}\over 2}+\Phi)},
\label{thss2}
\end{eqnarray}
where $\Phi({\bf r})$ is related to $f$ via the Poisson equation
(\ref{vh1}). The optimization problems (\ref{vh3}) or (\ref{vh4})
determine the most probable state of the $N$-body system (statistical
equilibrium state) resulting from a ``collisional'' relaxation. The
microcanonical ensemble (fixed energy) is appropriate to isolated
Hamiltonian systems like stellar systems. The relaxation towards
statistical equilibrium is driven by two-body encounters (Binney \&
Tremaine 1987) and the collisional evolution of stellar systems is
described by the gravitational Landau equation which increases the
Boltzmann entropy $S_B$ (H-theorem) at fixed mass $M$ and energy
$E$. Therefore, collisions (finite $N$ effects) eventually single out
the Boltzmann distribution among all possible stationary solutions of
the Vlasov equation. The canonical ensemble (fixed temperature) is
appropriate to dissipative systems in contact with a heat bath (of
non-gravitational origin) imposing its temperature, like in the model
of self-gravitating Brownian particles introduced by Chavanis et
al. (2002). The evolution of self-gravitating Brownian particles is
described by mean-field Fokker-Planck equations (Kramers and
Smoluchowski) which decrease the Boltzmann free energy $F_B$ at fixed
mass $M$ (Chavanis 2006b).

As is well-known, there is no global maximum of $S_{B}$ (resp. global
minimum of $F_{B}$) but there can exist {\it local} maxima of $S_B$
(resp. local minima of $F_B$) if the system is confined within a box
of radius $R$ and the energy (resp. temperature) is sufficiently
high. These {\it metastable} states are long-lived because their
lifetime scales as $e^{N}$ so that they are of prime interest in
astrophysics (Chavanis 2005a). The thermodynamical stability of the
system can be settled by using the turning point method of Poincar\'e
which was applied by Lynden-Bell \& Wood (1968) and Katz (1978) to
classical self-gravitating systems. Since we shall apply this method
in new situations (related to nonlinear dynamical stability problems),
we first describe how this method works in thermodynamics referring to
the previously quoted papers for technical details.

\begin{figure}[htbp]
\centerline{
\includegraphics[width=8cm,angle=0]{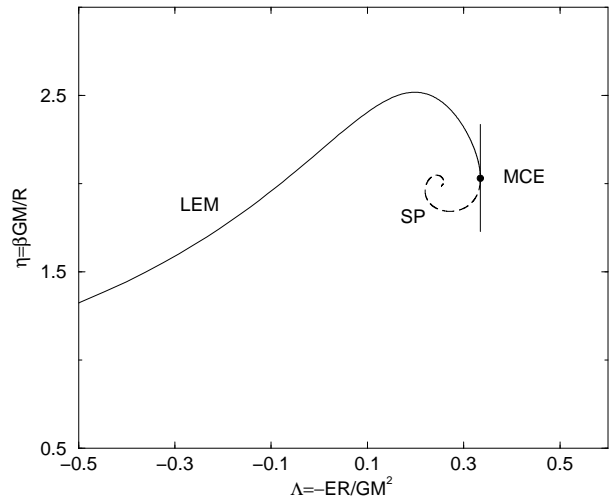}
} \caption[]{Series of equilibria of isothermal stellar systems
described by the Boltzmann entropy.  The relevant dimensionless energy
and dimensionless temperature are $\Lambda=-{ER\over GM^{2}}$ and
$\eta={\beta GMm\over R}$.  The thermodynamic limit $N\rightarrow
+\infty$ is such that these quantities remain fixed (Chavanis 2006b). In the
microcanonical ensemble, the stable states (local entropy maxima LEM)
are located on the full curve (microcanonical caloric curve). After
the turning point of energy MCE, the states become unstable (saddle
points SP of entropy).  }
\label{etalambdaMICRO}
\end{figure}

Suppose we want to maximize the Boltzmann entropy $S_{B}$ at fixed
mass $M$ and energy $E$ (microcanonical stability). The parameter
conjugate to the entropy $S_{B}$ with respect to the energy $E$ is the
inverse temperature $\beta=(\partial S_{B}/\partial
E)_{N,V}$. Therefore, if we plot the {\it series of equilibria} giving
$\beta$ as a function of $-E$, we have the following results:

(i) a change of stability can occur only at a turning point where $E$ is an extremum ($d\beta/dE$ infinite).

(ii) a mode of stability is lost if the curve rotates clockwise and gained if the curve rotates anti-clockwise.

The series of equilibria of isothermal stellar systems described by
the Boltzmann entropy is shown in Fig. \ref{etalambdaMICRO} and forms
a spiral. Now, we know that for $E$ sufficiently large,
the solutions are stable (maxima of $S_{B}$ at fixed $E$, $M$)
because, in this limit, self-gravity is negligible and the system
behaves like an ordinary gas in a box. From point (i), we conclude
that the whole upper branch until point MCE is stable and corresponds
to local entropy maxima (LEM). This stable part of the curve defines
the {\it microcanonical caloric curve}. At the first turning point of
entropy MCE, the curve rotates clockwise so that a mode of stability
is lost. As the curve spirals inward, more and more modes of stability
are lost (at each vertical tangent) so that the states past MCE are
unstable saddle points of entropy (SP).

\begin{figure}[htbp]
\centerline{
\includegraphics[width=8cm,angle=0]{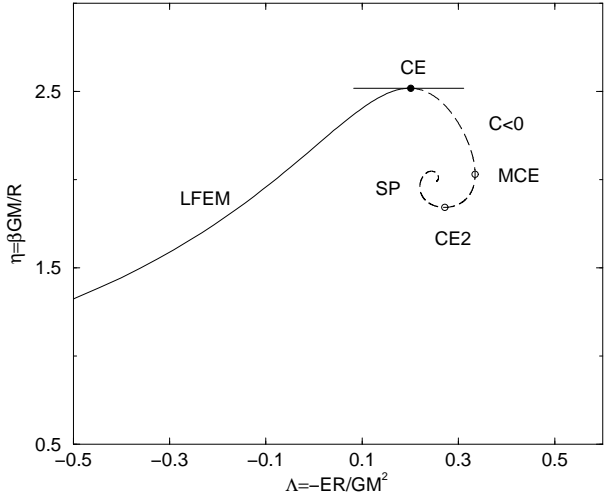}
} \caption[]{Series of equilibria of isothermal stellar systems
described by the Boltzmann free energy. In the canonical ensemble, the
stable states (local free energy minima LFEM) are located on the
full curve (canonical caloric curve). After the turning point of
temperature CE, the states become unstable (saddle points SP of
free energy). The region of negative specific heats between CE and MCE
is a domain of ensembles inequivalence: those states are stable in the microcanonical ensemble but unstable in the canonical ensemble. }
\label{etaLambdaCANO}
\end{figure}

Suppose now that we want to minimize the free energy $F$ at fixed
mass (canonical stability). This is equivalent to maximizing the
Massieu function $J=S-\beta E$ at fixed $M$. Now the parameter
conjugate to $J$ with respect to $\beta$ is $-E=(\partial J/\partial
\beta)_{N,V}$. Therefore, by using the turning point criterion
(properly adapted to the present case)\footnote{We just need to rotate
Fig. \ref{etaLambdaCANO} by $90^{o}$ and plot $E$ as a function of
$\beta$ to be in the conditions of application of (i) and (ii). } we
conclude that the upper branch of the spiral is stable in the
canonical ensemble until the point CE and corresponds to local minima
of free energy (FEM). This stable part of the curve defines the {\it
canonical caloric curve}. At the first turning point of temperature
CE, the curve rotates clockwise so that a mode of stability is
lost. As the curve spirals inward after CE, more and more modes of
stability are lost (at each horizontal tangent) so that the states are
saddle points of free energy (SP). We note that the states between CE
and MCE are stable in the microcanonical ensemble (local maxima of
entropy $S_{B}$ at fixed $E$, $M$) but unstable in the canonical
ensemble (saddle points of free energy $F_{B}$ at fixed $M$). Such
states have {\it negative specific heats} $C=(dE/dT)_{N,V}$. Negative
specific heats are forbidden in the canonical ensemble but they are
allowed in the microcanonical ensemble for systems with long-range
interactions for which the energy is non-additive (Thirring 1970,
Lynden-Bell \& Lynden-Bell 1977). Therefore, in this region, the
ensembles are {\it inequivalent}: the states with negative specific
heats cannot be reached by a canonical description (they are
unstable). Note that negative specific heats is a sufficient but not a
necessary condition of canonical instability: for example, the states
between MCE and CE2 have positive specific heats but they are
canonically (and microcanonically) unstable.

\subsection{Self-gravitating fermions}
\label{sec_fermi}

\begin{figure}[htbp]
\centerline{
\includegraphics[width=8cm,angle=0]{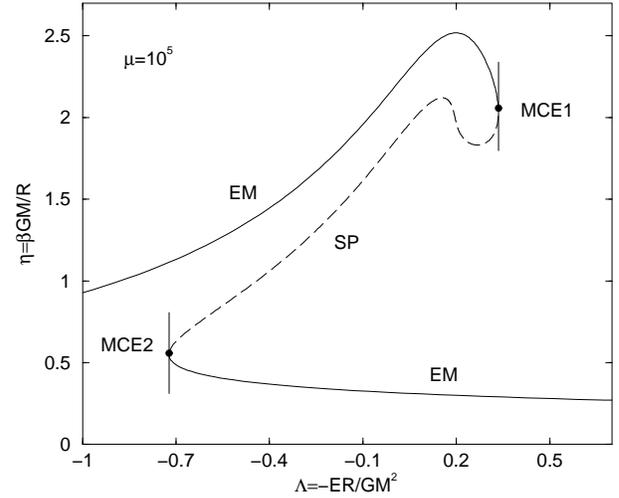}
} \caption[]{Series of equilibria of self-gravitating fermions
described by the Fermi-Dirac entropy for $\mu=10^5$. In the
microcanonical ensemble, the stable states (entropy maxima EM) are
located on the full curves. The upper branch corresponds to the
``gaseous phase'' and the lower branch to the ``condensed phase''. The
mode of stability lost at MCE1 is re-gained at MCE2. The dashed curve
corresponds to unstable saddle points of entropy. }
\label{etaLambda1e5}
\end{figure}

\begin{figure}[htbp]
\centerline{
\includegraphics[width=8cm,angle=0]{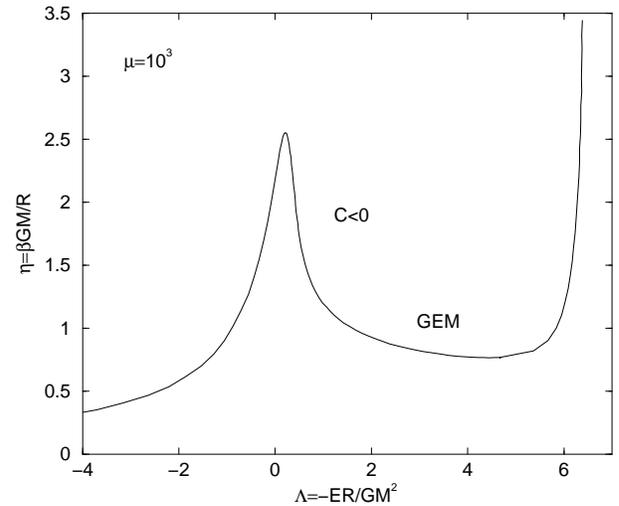}
} \caption[]{Series of equilibria of self-gravitating fermions
described by the Fermi-Dirac entropy for $\mu=10^3$. In the
microcanonical ensemble, the full series of equilibria is stable and
corresponds to global entropy maxima (GEM).  } \label{tata}
\end{figure}

\begin{figure}[htbp]
\centerline{
\includegraphics[width=8cm,angle=0]{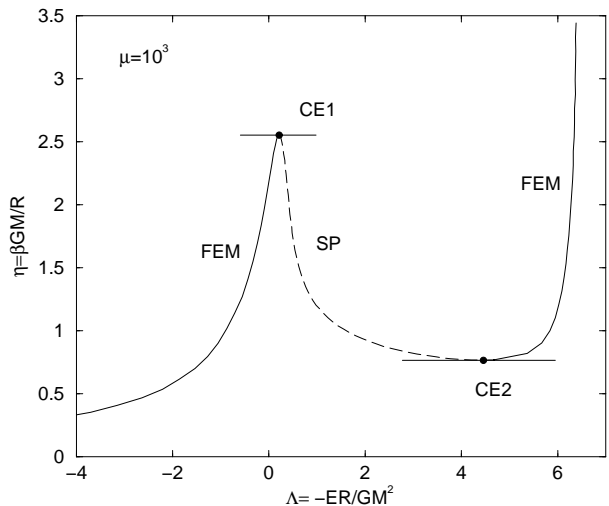}
} \caption[]{Series of equilibria of self-gravitating fermions
described by the Fermi-Dirac free energy for $\mu=10^3$. In the
canonical ensemble, the stable states (minima of free energy FEM) are
located on the full curves. The left branch corresponds to the ``gaseous
phase'' and the right branch to the ``condensed phase''. The mode of
stability lost at CE1 is re-gained at CE2. The dashed curve
corresponds to unstable saddle points of free energy. }
\label{Tcrit1000F}
\end{figure}

Another application of the turning point method where modes of
stability can be lost and then {\it re-gained} has been studied by Chavanis
(2002d) in the case of self-gravitating fermions. In that case, the
thermodynamical potential is the Fermi-Dirac entropy
\begin{eqnarray}
S=-{\eta_{0}\over m}\int\left\lbrace  {f\over\eta_{0}}\ln{f\over\eta_{0}}+\left (1- {f\over\eta_{0}} \right )\ln \left (1- {f\over\eta_{0}} \right )\right\rbrace d{\bf r}d{\bf v},\nonumber\\
\label{fd1}
\end{eqnarray}
and its extremization at fixed mass and energy leads to the Fermi-Dirac distribution
\begin{eqnarray}
f={\eta_{0}\over 1+\lambda e^{\beta m({v^{2}\over 2}+\Phi({\bf r}))}},
\label{fd2}
\end{eqnarray}
where $\lambda>0$ and $\eta_{0}=(m/h)^{3}$ is the maximum allowed
value of the distribution function due to the Pauli exclusion
principle. The form of the series of equilibria depends on the
degeneracy parameter $\mu\equiv \eta_{0}\sqrt{512 \pi^{4}G^{3}MR^{3}}$
\footnote{The degeneracy parameter can also be written
$\mu=17.259... (R/R_{*})^{3/2}$. It is proportional to the ratio, to
the power $3/2$, between the system size $R$ and the size $R_{*}$ of a
completely degenerate configuration (fermion ball at $T=0$) with mass
$M$ (Chavanis \& Rieutord 2003). Roughly speaking, it is a measure of
the importance of the small-scale regularization which is played here
by the Pauli exclusion principle. Thus, $\epsilon=1/\mu$ can be seen
as an effective small-scale cut-off. In terms of the Planck constant,
we have $1/\mu\sim \hbar^{3}$ so that the classical limit
($\hbar\rightarrow 0$) is recovered for $\mu\rightarrow
+\infty$.}. For a degeneracy parameter $\mu= 10^5$, the series of
equilibria is shown in Fig. \ref{etaLambda1e5}. The stability of the
configurations can be analyzed with the Poincar\'e turning point
argument. For large $E$, self-gravity is negligible with respect to
thermal motion and the system coincides with a classical perfect gas
in a box which is known to be thermodynamically stable.  Thus, from
(i) we conclude that the series of equilibria is microcanonically
stable until MCE1. At that point, the tangent is vertical and the
curve rotates clockwise so that a mode of stability is lost. However,
a second turning point of energy appears at MCE2 and the curve turns
anti-clockwise so that the mode of stability is re-gained.  Therefore,
the states between MCE1 and MCE2 are unstable saddle points of entropy
at fixed mass and energy (SP) while the states before MCE1 and after
MCE2 are maxima of entropy at fixed $E$, $M$.  These stable states
define the {\it physical} microcanonical caloric curve including the
metastable states (local entropy maxima). The transition between local
and global entropy maxima, the importance of metastable states and the
notion of phase transitions (first order and zeroth order) between a
gaseous phase and a condensed phase in the self-gravitating Fermi gas,
and more generally in self-gravitating systems with a small-scale
cut-off, are further discussed in Chavanis (2002d,2005a).

For a degeneracy parameter $\mu= 10^3$, the series of equilibria is
shown in Figs. \ref{tata} and \ref{Tcrit1000F}. Since there is no
turning point of energy, we conclude that the whole series of
equilibria is stable in the microcanonical ensemble and corresponds
to global entropy maxima (GEM) at fixed mass and energy. This forms
the microcanonical caloric curve (see Fig. \ref{tata}).
Alternatively, in the canonical ensemble (see Fig.
\ref{Tcrit1000F}), the presence of turning points of temperature and
the application of the Poincar\'e method (using the fact that for
high temperatures the system is stable) implies that the series of
equilibria is stable before CE1 and after CE2 (minima of free energy
at fixed mass) and unstable in between (saddle points of free energy
at fixed mass). The mode of stability lost at CE1 is re-gained at
CE2. The stable states on the solid curves  form the {\it physical}
canonical caloric curve including the metastable states. By
comparing Figs. \ref{tata} and \ref{Tcrit1000F} we note that the
region of negative specific heats $C<0$ in the microcanonical ensemble is
inaccessible (unstable) in the canonical ensemble where it is
replaced by a phase transition.

Note, finally, that the Fermi-Dirac distribution (\ref{fd2}) also
arises in the theory of violent relaxation for collisionless stellar
systems (Lynden-Bell 1967, Chavanis \& Sommeria 1998b). In that case,
the exclusion principle is of dynamical origin since it is related to
the conservation of the fine-grained DF by the Vlasov
equation. Furthermore, the maximum phase-space density $\eta_{0}$ is
related to the initial condition while $\eta_{0}\sim h^{-3}$ for
fermions. Degeneracy effects (in the sense of Lynden-Bell) may be
important in numerical simulations of violent relaxation (May \& van
Albada 1984) and, possibly, in the core of elliptical galaxies
(Stiavelli \& Bertin 1987, Chavanis \& Sommeria 1998b) and dark matter
(Kull et al. 1996).

\section{Dynamical stability  of stellar systems and barotropic stars}
\label{sec_thss}

\subsection{Thermodynamical analogy}
\label{sec_thanal}

Formally, the criteria of nonlinear dynamical stability (\ref{vh3})
and (\ref{vh4}) for ``collisionless'' stellar systems are {\it
similar} to criteria of thermodynamical stability for ``collisional''
self-gravitating systems (Sec. \ref{sec_thermo}) but they involve a more
general functional $S[f]=-\int C(f)d{\bf r}d{\bf v}$
 than the Boltzmann entropy $S_{B}[f]=-\int {f\over m}\ln
{f\over m}d{\bf r}d{\bf v}$.  Furthermore, they have a
completely different interpretation as they provide criteria of
nonlinear dynamical stability for a stationary solution of the Vlasov
equation while the maximization of the Boltzmann entropy at fixed mass
and energy provides a condition of thermodynamical stability (most
probable state) for the statistical equilibrium state of a $N$-body
stellar system reached for $t\rightarrow +\infty$.

Due to the formal resemblance between nonlinear dynamical stability
criteria and thermodynamical stability criteria, we can develop a
{\it thermodynamical analogy} to investigate the nonlinear dynamical
stability of collisionless stellar systems and use a vocabulary
borrowed from thermodynamics. In this analogy, $S$ is similar to an
entropy, $T=1/\beta$ is similar to a temperature and $F$ is similar
to a free energy (it is related to $S$ by a Legendre transform). The
criterion (\ref{vh3}) is similar to a condition of microcanonical
stability and the criterion (\ref{vh4}) is similar to a condition of
canonical stability. Since canonical stability implies
microcanonical stability (but not the converse), we recover the fact
that condition (\ref{vh8}) implies condition (\ref{vh7}). The
stability problems (\ref{vh3}) and (\ref{vh4}) can be studied by
plotting the linear series of equilibria $\beta(E)$, which is similar
to a caloric curve in thermodynamics.  Due to the above
thermodynamical analogy, the method of linear series of equilibria
can be used to settle the nonlinear dynamical stability of a
collisionless stellar system for a general functional $S$.  According
to the Poincar\'e theorem, a mode of stability is lost or gained at
a turning point. Note that the slope $dE/dT$ (where $\beta=1/T$ is
the Lagrange multiplier associated with the energy) plays an
important role because, according to the turning point argument, its
change of sign may be the signal of an instability. The slope
$C=dE/dT$ is similar to the specific heat in thermodynamics. We
shall give some illustrations of these results in the following
sections.

\subsection{The nonlinear Antonov first law}
\label{sec_anto}

For any stellar system with $f=f(\epsilon)$ and $f'(\epsilon)<0$,
there exists a corresponding barotropic star with the same equilibrium
density distribution. Indeed, defining the density and the pressure by
$\rho=\int fd{\bf v}=\rho(\Phi)$, $p={1\over 3}\int
fv^{2}d{\bf v}=p(\Phi)$, and eliminating the potential $\Phi$
between these two expressions, we find that $p=p(\rho)$. Writing
explicitly the density and the pressure in the form
\begin{eqnarray}
\rho=4\pi\int_{\Phi}^{+\infty}f(\epsilon)\sqrt{2(\epsilon-\Phi)}d\epsilon,
\label{anto1}
\end{eqnarray}
\begin{eqnarray}
p={4\pi\over 3}\int_{\Phi}^{+\infty}f(\epsilon)\lbrack 2(\epsilon-\Phi)\rbrack^{3/2}d\epsilon,
\label{anto2}
\end{eqnarray}
and taking the gradient of Eq. (\ref{anto2}), we obtain the condition
of hydrostatic equilibrium (\ref{ep5}).

\begin{figure}[htbp]
\centerline{
\includegraphics[width=8cm,angle=0]{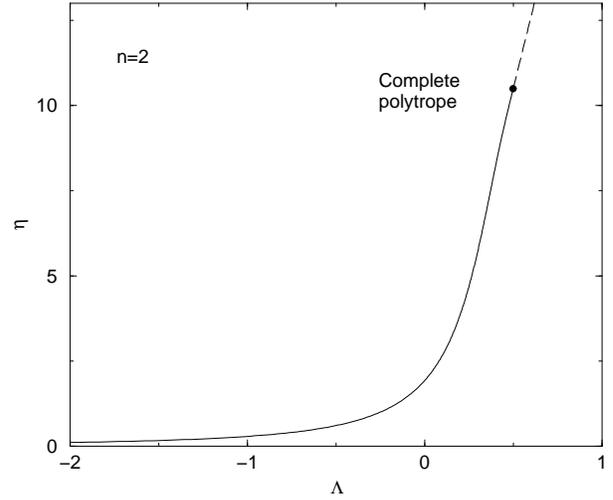}
} \caption[]{Series of equilibria for polytropic spheres with index
$n=2$ (see Chavanis 2002b and Chavanis \& Sire 2004 for the details of
the construction). In the thermodynamical analogy, this can be viewed
as an effective caloric curve giving the effective normalized inverse
temperature $\eta=M(G/n\Theta)^{n\over n-1}/R^{n-3\over n-1}$ (where
$n\Theta=K(1+n)/(4\pi)^{1/n}$ is a polytropic temperature proportional
to the polytropic constant $K$) as a function of the normalized energy
$\Lambda=-ER/GM^{2}$. The parameter $\eta$ is a monotonic function of
the Lagrange multiplier $\beta$ (through the relation
$K=\beta^{-(1-{3\over 2n})}$ obtained from Eqs. (24) and (31) of
Chavanis \& Sire 2004). Therefore, the curve $\eta-\Lambda$ is
equivalent, with more convenient variables, to the curve $\beta(E)$
issued from the optimization problems (\ref{vh3}) and (\ref{vh4}). The
first complete polytrope, with $\rho(R)=0$, is indicated by
$(\bullet)$. It corresponds to the terminal point in the series of
equilibria. The branch of complete polytropes with $R_*<R$ is
indicated by a dashed line. For $n<3$, the series of equilibria is
monotonic so that the ``ensembles'' are equivalent. Due to the
correspondence that we have found, this means that for $n<3$ both
stellar polytropes and polytropic stars are nonlinearly dynamically
stable with respect to the Vlasov-Poisson and Euler-Poisson systems
respectively. } \label{el2}
\end{figure}

As discussed in Sec. \ref{sec_ndss}, the minimization problem
(\ref{vh4}) provides a {\it sufficient} (but not necessary) condition
of nonlinear dynamical stability for a collisionless stellar
system. To solve this minimization problem, we can first minimize
$F[f]$ at fixed density $\rho({\bf r})$ to obtain $f_{*}({\bf r},{\bf
v})$. More precisely, $f_{*}$ is determined by
$C'(f_{*})=-\beta(v^2/2+\lambda({\bf r}))$ where $\lambda({\bf r})$ is
the Lagrange multiplier associated with the specification of the
density $\rho({\bf r})$. Since $\delta^2 F=(T/2)\int C''(f_*)(\delta
f)^2 d{\bf r}d{\bf v}>0$, $f_*$ really is a minimum of $F$ at fixed
$\rho({\bf r})$ (note that fixing $\rho({\bf r})$ automatically
determines the potential energy $W=(1/2)\int \rho\Phi d{\bf r}$ which
can thus be treated as a constant). From the optimal distribution
function $f_{*}$, we can compute the density $\rho=\int f_* d{\bf
v}=\rho(\lambda)$ and the pressure $p={1\over 3}\int f_* v^2 d{\bf
v}=p(\lambda)$. Eliminating $\lambda({\bf r})$ between these two
expressions, we obtain the equation of state of a barotropic gas
$p=p(\rho)$.  This is the same equation of state as in equilibrium
where $\lambda({\bf r})=\Phi({\bf r})+\alpha/\beta$ but the important
point is that the relation $p=p(\rho)$ also holds for perturbations
$\delta\rho$ around equilibrium. We can then express the functional
$F[f]$ as a functional of the density $\rho({\bf r})$ by setting
$F[\rho]=F[f_{*}]$. The calculations are detailed in Appendix \ref{sec_pass}
(see also Secs. 7.7 and 7.8 of Chavanis 2003a) and the functional
$F[\rho]$ can be finally written
\begin{eqnarray}
F[\rho]={1\over 2}\int\rho\Phi
d{\bf r}+\int\rho\int^{\rho}{p(\rho')\over
\rho'^{2}}d\rho'd{\bf r}, \label{anto4}
\end{eqnarray}
where $p(\rho)$ is the equation of state of the corresponding
barotropic gas defined above. We are thus led to consider the
minimization problem
\begin{eqnarray}
{\rm Min}\ \lbrace  F[\rho]\quad |\  M[\rho]=M \rbrace .
\label{anto3}
\end{eqnarray}
Now, we observe that the functional (\ref{anto4}) coincides with the
energy functional (\ref{ep4}) of a barotropic gas with ${\bf u}={\bf
0}$. We come therefore to the following conclusion: for a given
stellar system with $f=f(\epsilon)$ and $f'(\epsilon)<0$, if we know
that the corresponding barotropic star is nonlinearly dynamically
stable with respect to the Euler-Poisson system, then it is a minimum
of ${\cal W}[\rho,{\bf u}]$, hence of $F[\rho]$ (at fixed
mass). Therefore, the stellar system satisfies the criterion
(\ref{vh4}) so it is nonlinearly dynamically stable with respect to
the Vlasov-Poisson system. This leads to a nonlinear generalization of
the Antonov first law \footnote{This result has been obtained
independently in Chavanis (2003a) and Rein (2003,2005).}: ``a
spherical stellar system with $f=f(\epsilon)$ and $f'(\epsilon)<0$ is
nonlinearly dynamically stable with respect to the Vlasov-Poisson
system if the corresponding barotropic gas with the same equilibrium
density distribution is nonlinearly dynamically stable with respect to
the Euler-Poisson system''. However, the reciprocal is wrong in
general. A stellar system can be nonlinearly dynamically stable
according to the ``microcanonical'' criterion (\ref{vh3}) while it
does not satisfy the ``canonical'' criterion (\ref{vh4}) so that the
corresponding barotropic star is dynamically unstable. Due to the
thermodynamical analogy, the nonlinear Antonov first law for
collisionless stellar systems has the same status as the fact that:
``canonical stability implies microcanonical stability in
thermodynamics''. The converse is wrong in case of ``ensembles
inequivalence'' which is generic for systems with long-range
interaction such as gravity. To the point of view of their nonlinear
dynamical stability, the crucial difference between stars and galaxies
is that, for galaxies, both $S$ and $E$ are {\it individually}
conserved by the Vlasov equation while in the case of stars only the
total energy ${\cal W}$ is conserved by the Euler
equations. Therefore, the most refined stability criterion for
spherical galaxies is (\ref{vh3}) while the most refined stability
criterion for stars is (\ref{ej4}), which is included in
(\ref{vh3}). This is the intrinsic reason why a spherical galaxy can
be stable even if the corresponding barotropic star is unstable.
Therefore, the set of stable spherical galaxies is larger than the set
of stable barotropic stars. Similarly, in thermodynamics, the set of
stable microcanonical equilibria is wider than the set of stable
canonical equilibria (it includes in particular states with negative
specific heats) because there is an additional constraint in the
problem: the energy.

\subsection{Application to polytropes}
\label{sec_poly}

\begin{figure}[htbp]
\centerline{
\includegraphics[width=8cm,angle=0]{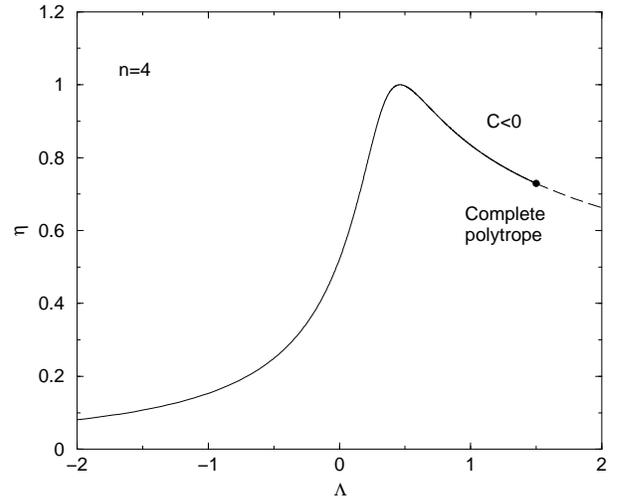}
} \caption[]{Series of equilibria for polytropic spheres with index
$n=4$. The first complete polytrope, with $R_{*}=R$, is indicated by
$(\bullet)$ and the branch of complete polytropes with $R_*<R$ by a
dashed line.  Since they lie after the turning point of
``temperature'' (appearing for $n=3$) and before the turning point of
energy (appearing for $n=5$), in the thermodynamical analogy, they are
unstable in the ``canonical'' ensemble (saddle points of ``free
energy'' $F$ at fixed $M$) but stable in the ``microcanonical''
ensemble (maxima of ``entropy'' $S$ at fixed $E$,$M$). Due to the
correspondance that we have found (see Sec. \ref{sec_anto}), these
effective thermodynamical stability criteria mean in reality that
stellar polytropes with $n=4$ are nonlinearly dynamically stable
stationary solutions of the Vlasov-Poisson system but polytropic stars
with $n=4$ are linearly dynamically unstable stationary solutions of
the Euler-Poisson system. This is similar to a situation of ``ensembles
inequivalence'' in thermodynamics. In particular, complete polytropes
with $n=4$ lie in a region of effective negative specific heats
$C<0$. } \label{el4}
\end{figure}

\begin{figure}[htbp]
\centerline{
\includegraphics[width=8cm,angle=0]{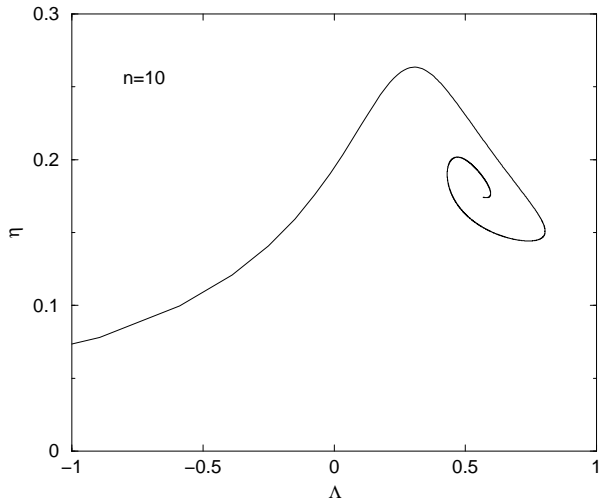}
} \caption[]{Series of equilibria for polytropic spheres with index
$n=10$. For $n>5$, self-gravitating polytropic spheres are unbounded
and have infinite mass. Thus, equilibrium distributions only exist if
the system is confined within a box. Box-confined gaseous polytropes
cease to be stable with respect to the Euler-Poisson system after the
first turning point of ``temperature'' $\eta$ and box-confined stellar
polytropes cease to be stable with respect to the Vlasov-Poisson
system after the first turning point of energy $\Lambda$. These
stability criteria can be expressed in terms of the density contrast
between center and edge which parameterizes the series of equilibria,
like for isothermal spheres. } \label{el10}
\end{figure}

It is interesting to consider the application of these results to
the case of polytropes. Stellar polytropes extremize the Casimir functional
\begin{eqnarray}
S_{q}=-{1\over q-1}\int (f^{q}-f)d{\bf r}d{\bf v}, \label{caspoly}
\end{eqnarray}
at fixed mass and energy. This leads to the class of distribution functions
\begin{eqnarray}
f=\biggl \lbrack \mu-{\beta (q-1)\over q}
\epsilon\biggr \rbrack_{+}^{1\over q-1},
\label{polydf}
\end{eqnarray}
with the convention that $f=0$ when the term in brackets becomes
negative.  The usual polytropic index $n$ is related to the parameter
$q$ by $n={3\over 2}+{1\over q-1}$.  We can now plot the series of
equilibria $\beta(E)$ issued from the optimization problems
(\ref{vh3}) and (\ref{vh4}) with the polytropic functional
(\ref{caspoly}). The equations determining these curves (for various
index $n$) have been given in previous papers (Chavanis 2003a,
Chavanis \& Sire 2004) so we shall only describe the results and
present a ``gallery'' of caloric curves.  We recall that the system is
confined within a box of radius $R$ so that the series of equilibria
is essentially formed by {\it incomplete polytropes} (represented in
solid line) which are held by the box so that $\rho(R)>0$. The series
of equilibria can be parameterized by the density contrast ${\cal
R}=\rho(0)/\rho(R)$ between center and edge.  The terminal point of
the series of equilibria corresponds to the {\it first complete
polytrope} (represented by a bullet) for which the density drops to
zero precisely at the box radius $\rho(R)=0$ (if $n<5$). In the
present context, this is the most important point in the series of
equilibria because it connects the branch of incomplete polytropes to
the branch of {\it complete polytropes} (represented by a dashed line)
for which the density drops to zero at $R_*<R$ (if $n<5$). These are
the physical structures in astrophysics since they are not affected by
an artificial box. We thus have to investigate the stability of the
first complete polytrope (bullet) to determine the stability of all
the complete polytropes \footnote{In fact, complete polytropes are
equivalent by the homology theorem (Chandrasekhar 1942) but this is
not the case in general for other stellar systems.} which are of
physical relevance. To that purpose, we just need to investigate the
presence of turning points in the series of equilibria. Therefore, our
method is very simple and very general. It just requires to solve the
{\it first order variations} of the optimization problems (\ref{vh3})
and (\ref{vh4}) and use the turning point argument to obtain powerful
results of nonlinear dynamical stability for barotropic stars and
spherical galaxies.

For $n<3$ the series of equilibria $\beta(E)$ for polytropes is
monotonic (see Fig. \ref{el2}) so that, according to the Poincar\'e
criterion, all the polytropes of the series are maxima of $S$ (at
fixed $E$, $M$) and minima of $F$ (at fixed $M$). In our
thermodynamical analogy, this corresponds to a situation of ensemble
equivalence. These results imply that, for $n<3$, both gaseous
polytropes and stellar polytropes are nonlinearly dynamically stable
with respect to the Euler-Poisson and Vlasov-Poisson systems
respectively. It can be shown that a turning point of ``temperature''
$\beta$ appears for $n>3$ (Chavanis 2002b). Therefore, complete
polytropes (and incomplete polytropes with high density contrast lying
after the turning point of ``temperature'') cease to be minima of $F$
(at fixed mass) for $n>3$ (see Fig. \ref{el4}). In that case, they lie
in a region of the series of equilibria where the slope $dE/dT$ is
negative. In the thermodynamical analogy, this is similar to a loss of
canonical stability in a region of negative specific heats. Using the
preceding results, we conclude that complete polytropic stars are
nonlinearly dynamically stable with respect to the Euler-Poisson
system for $n<3$ but they become linearly dynamically unstable for
$n>3$. On the other hand, a turning point of energy $E$ appears for
$n>5$ (see Fig. \ref{el10}). Therefore, complete polytropes are maxima
of $S$ (at fixed mass and energy) for $n<5$. Polytropes cease to be
self-confined for $n>5$ and incomplete polytropes with high density
contrast lying after the turning point of energy become unstable with
respect to the Vlasov-Poisson system for $n>5$. This is similar to a
loss of microcanonical stability in the thermodynamical analogy. In
conclusion, stellar polytropes are nonlinearly dynamically stable with
respect to the Vlasov-Poisson system for $3/2<n<5$ (they do not exist
for $n<3/2$) and they become dynamically unstable for $n>5$ above a
critical density contrast. On the other hand, gaseous polytropes are
nonlinearly dynamically stable with respect to the Euler-Poisson
system for $0<n<3$ and they become dynamically unstable for $n>3$
above a critical density contrast. For $3<n<5$, complete stellar
polytropes are nonlinearly dynamically stable with respect to the
Vlasov-Poisson system while corresponding polytropic stars are
dynamically unstable with respect to the Euler-Poisson system. This is
similar to a situation of ensembles inequivalence, in a region of
negative specific heats, in thermodynamics.  Of course, the dynamical
stability of stellar polytropes and polytropic stars has been
established for a long time in astrophysics but our method relying on
the appearance of turning points in the series of equilibria and the
interpretation of these results, and of the Antonov first law, in
terms of a thermodynamical analogy and a situation of ensembles
inequivalence is new and original. Furthermore, only linear dynamical
stability is considered in general while the criteria (\ref{vh3}) and
(\ref{vh4}) are criteria of {\it nonlinear} dynamical stability.

\subsection{The main point: correspondances}
\label{sec_main}

In summary, we have the following analogy. In thermodynamics, if we
plot the series of equilibria $\beta(E)$ we have the following result:

$\bullet$ After a turning point of temperature (or dimensionless
parameter $\eta$), the system becomes canonically unstable (saddle
point of free energy $F$ at fixed $M$). After a turning point of
energy (or dimensionless parameter $\Lambda$), the system becomes
microcanonically unstable (saddle point of entropy $S$ at fixed $E$,
$M$).  If the series of equilibria does not unwind, the system
remains thermodynamically unstable after these turning points. Before
these turning points, the system is thermodynamically stable.

Concerning the nonlinear dynamical stability of collisionless stellar
systems with $f=f(\epsilon)$, $f'(\epsilon)<0$ and corresponding
barotropic stars with the same density distribution, if we plot
the series of equilibria $\beta(E)$ we have the following result:

$\bullet$ After a turning point of ``temperature'' $\beta$ (or
dimensionless parameter $\eta$), the barotropic gas becomes
dynamically unstable with respect to the Euler-Poisson system. After a
turning point of energy $E$ (or dimensionless parameter $\Lambda$),
the stellar system becomes dynamically unstable with respect to the
Vlasov-Poisson system. If the series of equilibria does not unwind,
the system remains unstable after these turning points. Before these
turning points, the system is nonlinearly dynamically stable.

Schematically, we have the correspondances:
\begin{eqnarray}
(i)\  Nonlinear \ dynamical \ stability \ of \nonumber\\
barotropic \ stars \leftrightarrow ``canonical'' \ criterion\nonumber\\
(ii) \ Nonlinear \ dynamical \ stability \ of \ spherical  \nonumber\\
stellar \ systems
\leftrightarrow ``microcanonical'' \ criterion\nonumber\\
(iii)\  Nonlinear \ Antonov \ first \ law  \nonumber\\
\leftrightarrow canonical \ stability \ implies  \ microcanonical \ stability \nonumber\\
\nonumber
\end{eqnarray}

\subsection{Other dimensions of space}
\label{sec_otherdim}

For completeness, we also show the series of equilibria for
polytropic spheres in other dimensions of space $d$. These results can
be of interest for mathematicians (see, e.g., Lemou et al. 2005). Furthermore,
the problem is rich and interesting because it exhibits several
critical dimensions with intriguing consequences concerning the
particular dimension of our universe (see Chavanis 2004).  The
construction of these curves is detailed in Chavanis \& Sire (2004)
and some new curves are reported here in order to have a complete
picture of the problem in parameter space $(d,n)$. We first recall
that physical stellar polytropes (whose DF is a {\it decreasing}
function of the energy) only exist for $n\ge n_{3/2}\equiv d/2$ while
gaseous polytropes can be considered a priori for any index (but we
shall mostly consider the case $n\ge 0$ here). Furthermore, complete
polytropes exist for all index if $d\le 2$ and for $n<n_{5}\equiv
(d+2)/(d-2)$ if $d>2$. Here, we only describe the stability of
complete polytropes, i.e. the terminal point (bullet) in the series of
equilibria.

\begin{figure}[htbp]
\centerline{
\includegraphics[width=8cm,angle=0]{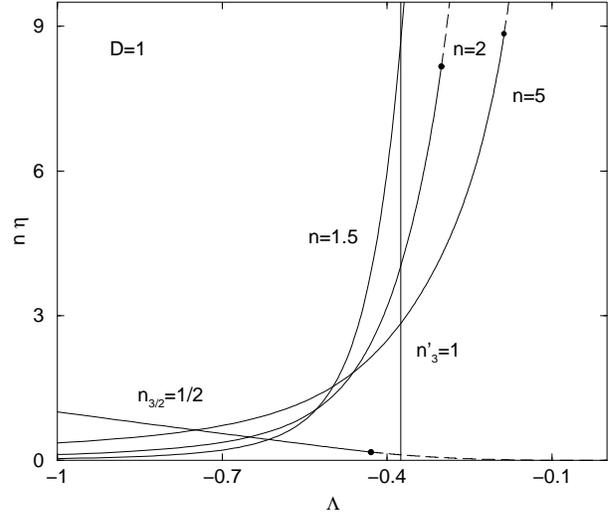}
} \caption[]{Series of equilibria for polytropic spheres in $d=1$.}
\label{elD1P}
\end{figure}

\begin{figure}[htbp]
\centerline{
\includegraphics[width=8cm,angle=0]{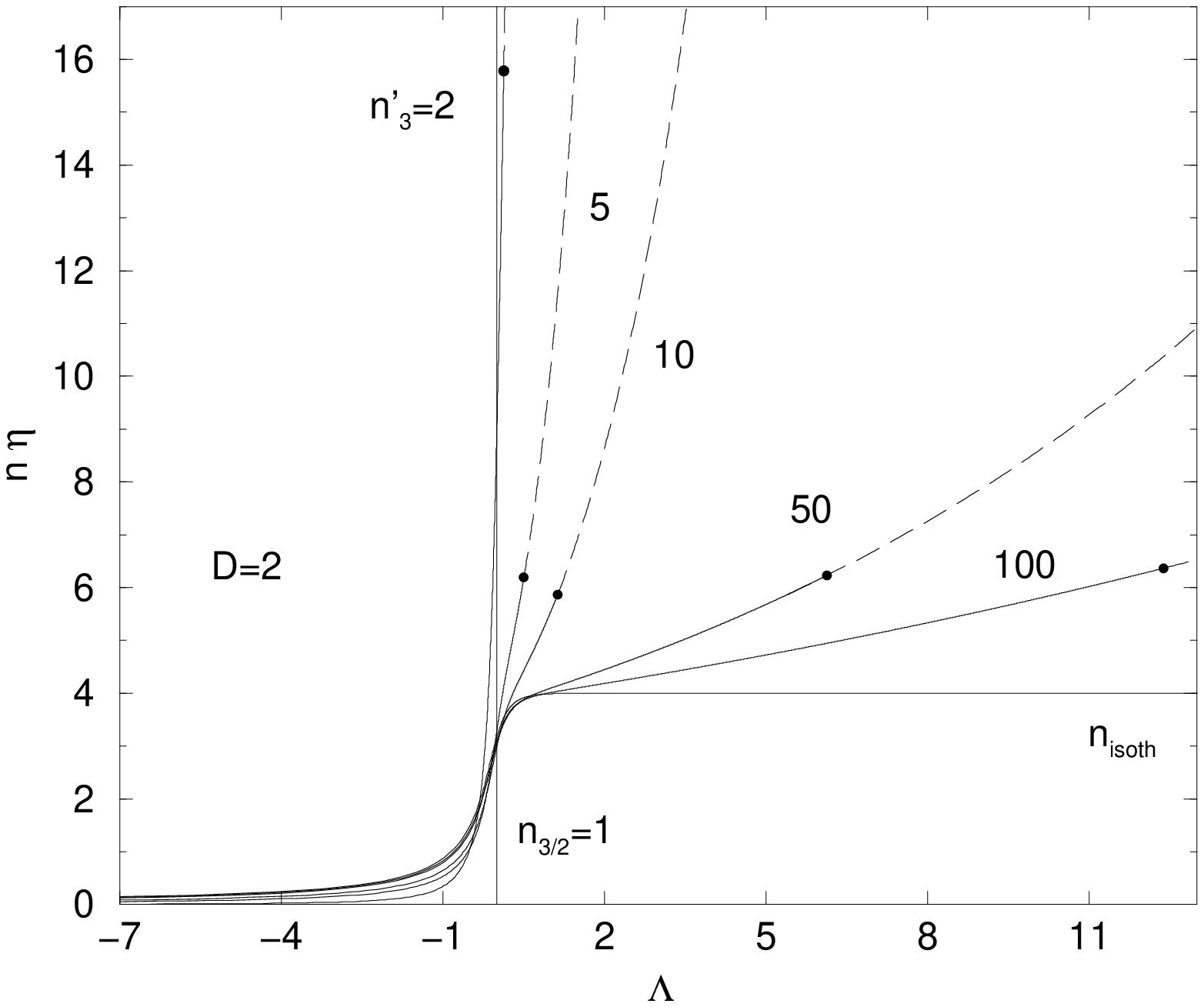}
} \caption[]{Series of equilibria for polytropic spheres in $d=2$.}
\label{elD2P}
\end{figure}

For $d=1$ and $d=2$ the series of equilibria are monotonic (see
Figs. \ref{elD1P}-\ref{elD2P}) implying that both stellar polytropes
with $n\ge n_{3/2}$ and polytropic stars with $n\ge 0$ are nonlinearly
dynamically stable (the ``ensembles'' are equivalent).

\begin{figure}[htbp]
\centerline{
\includegraphics[width=8cm,angle=0]{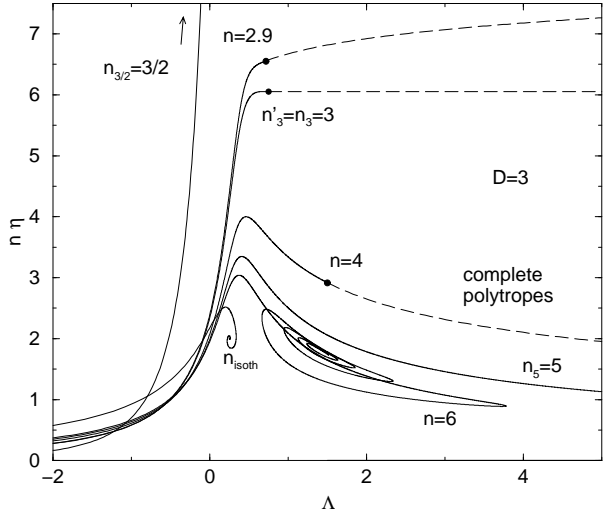}
} \caption[]{Series of equilibria for polytropic spheres in $d=3$.}
\label{elD3}
\end{figure}

For $2<d<4$, complete stellar polytropes exist for $n_{3/2}<n<n_{5}$
and they are nonlinearly dynamically stable (the turning point of
energy appears for $n\ge n_{5}$).  On the other hand, complete
polytropic stars are nonlinearly dynamically stable for
$0\le n<n_{3}\equiv d/(d-2)$ ($n_{3}$ is the index at which the turning
point of ``temperature'' appears) and unstable for $n_{3}\le n<n_{5}$
(see Figs. \ref{elD3}).

\begin{figure}[htbp]
\centerline{
\includegraphics[width=8cm,angle=0]{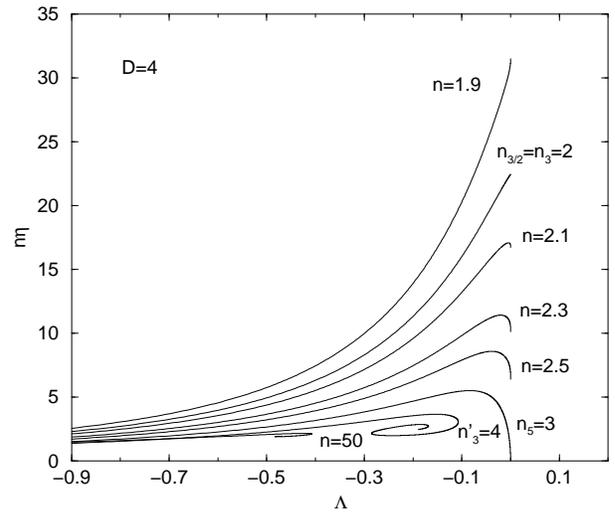}
} \caption[]{Series of equilibria for polytropic spheres in $d=4$.}
\label{elD4}
\end{figure}

For $d=4$, complete stellar polytropes exist
for $n_{3/2}=2<n<n_{5}=3$ and they all have energy $E=0$ (see Fig.
\ref{elD4}) as can be deduced from the results of Chavanis \& Sire (2004). The series of equilibria has a vertical tangent at
$E=0$ implying that complete stellar polytropes are marginally stable
in $d=4$ (they lie precisely at the turning point of energy). This is
consistent with the Virial theorem in $d=4$ for an isolated system
(see Appendix C of Chavanis \& Sire 2005). On the other hand, complete
polytropic stars are nonlinearly dynamically stable for $0\le
n<n_{3}=2$ and unstable for $n_{3}=2\le n<n_{5}=3$ due to the presence of
the turning point of ``temperature''.

For $d\ge 5$, there are no complete stellar
polytropes with $n\ge n_{3/2}$ (the cross-over between $n_{5}$ and
$n_{3/2}$ takes place for $d=2(1+\sqrt{2})\simeq 4.83...$). On the
other hand, complete polytropic stars are nonlinearly dynamically
stable for $n<n_{3}$ and unstable for $n_{3}\le n<n_{5}$ due to
the presence of the turning point of ``temperature'' (see Fig.
\ref{elD5}). For $n=n_{3/2}$ the series of equilibria makes angular
points as explained in Appendix C of Chavanis \& Sire (2004). For
$n>n_{5}$, the series of equilibria forms a spiral whose sense of
rotation changes for $n=n_{3/2}$. For $d\ge 10$ the situation
slightly changes in the sense that the spiral shrinks to a point
after an index $n_{-}$ (see Chavanis \& Sire 2004).  It is of
interest to note that the spatial dimension $d=10$ is special in classical
gravity (see Sire \& Chavanis 2002) as also found in theories of
supergravity (e.g. Cariglia \& Mac Conamhna 2005). We do not know
whether this is a coincidence of if it bears more physical meaning.

\begin{figure}[htbp]
\centerline{
\includegraphics[width=8cm,angle=0]{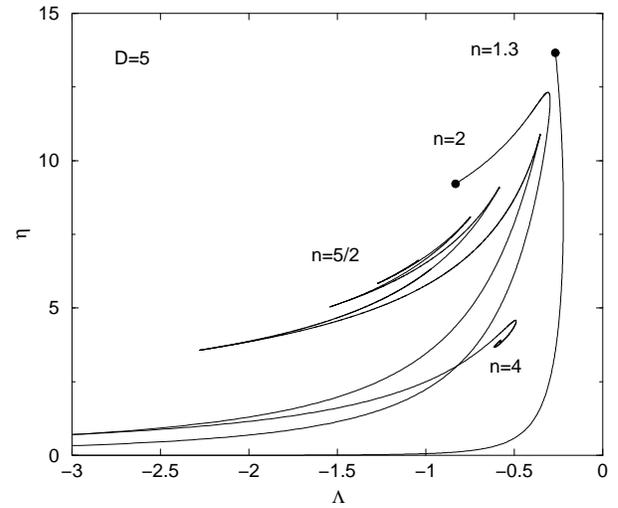}
} \caption[]{Series of equilibria for polytropic spheres in $d=5$.
For clarity, the curves have been multiplied by some factor to enter
in the same figure.} \label{elD5}
\end{figure}

We note finally that classical white dwarf stars are equivalent to
polytropes with index $n_{3/2}=d/2$ (Chavanis \& Sire
2004). Therefore, they are self-confined (complete) if
$n_{3/2}<n_{5}$, i.e. $d<2(1+\sqrt{2})\simeq 4.83...$ and they are
stable if $n_{3/2}<n_{3}$, i.e. $d<4$. For $d\ge 4$, quantum mechanics
(Pauli exclusion principle) cannot stabilize matter against
gravitational collapse (Chavanis 2004). Relativistic white dwarf stars
are equivalent to polytropes with index $n'_{3}=d$. Therefore, they
are self-confined (complete) if $n'_{3}<n_{5}$, i.e. $d<{1\over
2}(3+\sqrt{17})=3.56...$ and they are stable if $n'_{3}<n_{3}$,
i.e. $d<3$. These results will be discussed more specifically in
another paper (Chavanis 2006c).

\section{Phenomenology of violent relaxation} \label{sec_phen}

In this section we show that the criterion of nonlinear dynamical
stability (\ref{vh3}) is remarkably consistent with the phenomenology
of violent relaxation. Collisionless stellar systems like elliptical
galaxies are believed to have achieved a steady state as a result of
phase mixing and {\it incomplete violent relaxation} (Lynden-Bell
1967). During this process, the fine-grained distribution function
$f({\bf r},{\bf v},t)$ develops intermingled filaments at smaller and
smaller scales so that it does not converge toward any steady
distribution.  However, if we locally average over these filaments,
the coarse-grained distribution function $\overline{f}({\bf r},{\bf
v},t)$ is expected to converge towards a steady state
$\overline{f}({\bf r},{\bf v})$ which is a stable stationary solution
of the Vlasov equation. This is referred to as {\it weak convergence}
in mathematics.  Thus, the theory of violent relaxation explains {\it
how} a collisionless stellar system, initially out-of-equilibrium, can
reach a steady state of the Vlasov equation (on a coarse-grained
scale) due to mean-field effects. Since this metaequilibrium state
results from a complicated mixing process it is expected to be
particularly robust and have nonlinear dynamical stability properties.

Now, Tremaine et al. (1986) have shown that the functionals
$S[\overline{f}]=-\int C(\overline{f})d{\bf r}d{\bf v}$, where $C$ is
convex, calculated with the coarse-grained distribution function
increase during violent relaxation in the sense that $S(t)\ge S(0)$
for $t\ge 0$ where it is assumed that the DF is not mixed at $t=0$,
i.e. $\overline{f}({\bf r},{\bf v},0)=f({\bf r},{\bf v},0)$. For this
reason, these functionals are called generalized $H$-functions.  By
contrast, the energy $E[\overline{f}]$ and the mass $M[\overline{f}]$
calculated with the coarse-grained DF are approximately conserved. The
fact that $S[\overline{f}]$ increases at fixed $E[\overline{f}]$ and
$M[\overline{f}]$ is similar to a generalized selective decay
principle (for $-S$), like in 2D turbulence, which is due to phase
mixing and coarse-graining. Phenomenologically, this {\it may} suggest
that the system will reach a steady state $\overline{f}({\bf r},{\bf
v})$ which maximizes a certain $H$-function (non-universal) at fixed
mass and energy. This is consistent with the criterion of nonlinear
dynamical stability (\ref{vh3}) provided that we interpret the
distribution function as the {\it coarse-grained} distribution
function. Phase mixing and coarse-graining explain why we have to {\it
maximize} a certain Casimir functional although the evolution is
dissipationless and the Casimirs are conserved on the fine-grained
scale \footnote{The physical reason is the following. During mixing
$D\overline{f}/Dt\neq 0$ and the $H$-functions $S[\overline{f}]$
increase. Once it has mixed, $D\overline{f}/Dt=0$ so that $\dot
S[\overline{f}]=0$ meaning that the Casimirs of the coarse-grained DF are
now conserved. If $\overline{f}({\bf r},{\bf v},t)$ has been brought to a
maximum $\overline{f}_{0}({\bf r},{\bf v})$ of a certain $H$-function
(during violent relaxation) and since $S[\overline{f}]$ is conserved
(after mixing), then $\overline{f}_{0}$ is a nonlinearly dynamically
stable steady state of the Vlasov equation against coarse-grained
perturbations, according to the criterion (\ref{vh3}).}.

In reality, the problem is more complicated than that because (i) the
$H$-functions do not increase necessarily monotonically with time
(they are not Lyapunov functionals satisfying $\dot S\ge 0$) (ii) the
system can converge toward a stationary solution of the Vlasov
equation which does not depend on the energy $\epsilon$ alone, and
thus which does not maximize an $H$-function at fixed $E$, $M$. For
example, the velocity distribution of stars in elliptical galaxies is
anisotropic and depends on the angular momentum $J=|{\bf r}\times {\bf
v}|$ in addition to energy $\epsilon$. Furthermore, real stellar
systems are in general not spherically symmetric so their DF does not
only depend on $\epsilon$ and $J$.  Therefore, more general stationary
solutions of the Vlasov equation must be constructed in consistency
with the Jeans theorem (Binney \& Tremaine 1987). However, their
dynamical stability is more complicated to investigate.

More than constructing particular stationary solutions of the Vlasov
equation in order to reproduce observations, it is a challenging issue
to {\it predict} the distribution function of a galaxy resulting from
incomplete violent relaxation. Lynden-Bell (1967) first addressed this
question by developing a {\it statistical} theory of violent
relaxation. He predicted a DF of the form
$\overline{f}_{LB}=\overline{f}_{LB}(\epsilon)$ with
$\overline{f}'_{LB}(\epsilon)<0$ which maximizes a certain
$H$-function at fixed $E$, $M$ whose form depends on the initial
conditions through the Casimirs (see Chavanis 2006a). In a dilute
approximation which seems to apply to elliptical galaxies, the DF
predicted by Lynden-Bell reduces to the Maxwell-Boltzmann distribution
$f\sim e^{-\beta\epsilon}$ which extremizes the Boltzmann entropy at
fixed mass and energy. However, due to incomplete relaxation, this
{\it most mixed state} is not reached in practice \footnote{This is
obvious in the case of self-gravitating systems because the isothermal
distribution coupled to the Poisson equation has infinite mass so it
is ill-behaved mathematically. However, violent relaxation is in
general incomplete also in systems for which the isothermal
distribution is well-behaved mathematically. This is the case, e.g.,
in 2D turbulence (Chavanis \& Sommeria 1998a, Brands et al. 1999).} 
and other stationary solutions of the Vlasov equation can be reached
instead. For example, ellipticals are well relaxed (in the sense of
Lynden-Bell) in their inner region (leading to an isotropic isothermal
core with density profile $\sim r^{-2}$) while they possess radially
anisotropic envelopes (with density profile $\sim r^{-4}$). One
success of the Lynden-Bell theory is to explain the isothermal cores
of ellipticals without recourse to collisions.

The process of incomplete relaxation, corresponding to a lack of
ergodicity (partial mixing), could be taken into account by developing
{\it dynamical} models of violent relaxation as proposed in Chavanis
et al. (1996) and Chavanis (1998) to understand what limits
relaxation.  Alternatively, in a series of  papers, Bertin,
Stiavelli and Trenti have proposed other ideas to account for
incomplete violent relaxation. Stiavelli \& Bertin (1987) introduced an $f^{(\infty)}$
model of the form:
\begin{eqnarray}
f^{(\infty)}=A(-\epsilon)^{3/2}e^{-a\epsilon-cJ^{2}/2}, \qquad (\epsilon\le 0)
\label{sbinf}
\end{eqnarray}
and $f=0$ for $\epsilon>0$, based on the possibility that the {\it a
priori} probabilities of microstates are not equal due to kinetic
constraints. This model reproduces many properties of ellipticals but
it has the undesired feature of being ``too isotropic''. Then, they
introduced another model based on a modification of the Lynden-Bell
statistical theory. They considered the maximization of the Boltzmann
entropy (in Lynden-Bell's sense) at fixed mass, energy
{\it and} a third global quantity $Q=\int J^{\nu}|\epsilon|^{-3\nu/4}f
d{\bf r}d{\bf v}$ which is argued to be approximately conserved
during violent relaxation (Trenti et al. (2005) have checked that $Q$ is
reasonably well-conserved in numerical experiments of dissipationless
galaxy formation).  This variational principle results in a family of
$f^{(\nu)}$ models:
\begin{eqnarray}
f^{(\nu)}=A\ {\rm exp}\left\lbrack -a\epsilon-d\left ({J^{2}\over |\epsilon|^{3/2}}\right )^{\nu/2}\right\rbrack, \qquad (\epsilon\le 0)
\label{sbnu}
\end{eqnarray}
and $f=0$ for $\epsilon>0$. These models are able to fit products of
$N$-body simulations over nine orders of magnitude in density and to
reproduce the de Vaucouleur's $R^{1/4}$ law (or more general $R^{1/n}$
laws) of ellipticals (Trenti et al. 2005).  The introduction of
additional constraints in the variational principle could be a way to
take into account effects of incomplete violent relaxation.  Bertin \&
Trenti (2003) also showed that these models exhibit a sort of
``gravothermal catastrophe'' for collisionless stellar systems. They
argue that this corresponds to a thermodynamical instability where
internal collisionality is due not to stellar encounters (which are
negligible because of the very long relaxation time) but to
several dynamical causes making some relaxation proceed even at
current epochs. According to the discussion of Sec. \ref{sec_thanal},
the maximization of $S$ at fixed $E$, $M$ and $Q$ could also be viewed
as a condition of nonlinear dynamical stability for the anisotropic DF
(\ref{sbnu}) with respect to the Vlasov-Poisson system.

\section{Conclusion}
\label{sec_conclusion}

In this paper, we have shown that the dynamical (collisionless
evolution) and the thermodynamical (collisional evolution) stability
of stellar systems can be analyzed by similar methods. The
thermodynamical stability of stellar systems has been investigated
using optimization methods based on the maximization of the Boltzmann
entropy at fixed mass and energy (Antonov
1962, Lynden-Bell \& Wood 1968, Katz 1978, Padmanabhan 1990, Chavanis
2002a). In contrast, the dynamical stability of collisionless stellar
systems described by the Vlasov-Poisson system has been customarily
analyzed by linearizing the Vlasov equation around a steady state and
studying the resulting eigenvalue equation (Binney \& Tremaine
1987). We have approached the dynamical stability problem by a
different method, through the maximization of a Casimir functional (or
$H$-function) at fixed mass and energy.  This provides a condition of
{\it nonlinear dynamical stability}.  Furthermore, this allows us to
develop a close formal analogy with the thermodynamical problem. Due
to this analogy, we can use methods similar to those introduced in
thermodynamics (series of equilibria, Legendre transforms,
Poincar\'e's turning point argument,...) to settle the nonlinear
dynamical stability of a spherical stellar system. Thus, the existence
of turning points of temperature and energy which respectively mark
the onset of canonical and microcanonical instability in
thermodynamics also mark the onset of instability for
barotropic stars and spherical stellar systems in the dynamical
stability problem with respect to the Euler and Vlasov equations. For
example, when we plot the series of equilibria of polytropes, a
turning point of ``temperature'' appears for $n\ge 3$ and a turning
point of energy appears for $n\ge 5$; these are precisely the indices
marking the dynamical instability thresholds of polytropic stars and
stellar polytropes respectively. These indices had been previously
found by other authors using very different arguments (see Binney \&
Tremaine 1987). In this paper, we have shown that the nonlinear
Antonov first law could be viewed as a manifestation of ``ensembles
inequivalence'' in self-gravitating systems. This is an original
interpretation of this law. This {\it thermodynamical analogy} allows us
to unify two types of studies (dynamical and thermodynamical) that had
been previously approached by very different methods.

There are, however, some limitations in our approach: (i) Our criteria
of nonlinear dynamical stability only apply to {\it spherical stellar
systems} described by a DF of the form $f=f(\epsilon)$ with
$f'(\epsilon)<0$. It is not clear how they can be extended to more
general steady distributions of the Vlasov equation satisfying the
Jeans theorem. However, in the case of anisotropic stellar systems, we
have proposed (see Chavanis 2003a) considering the maximization of an
$H$-function at fixed mass, energy {\it and} ${\bf L}_{2}=\int f({\bf
r}\times {\bf v})^{2}d{\bf r}d{\bf v}$ which could be viewed as an
adiabatic invariant (it is not strictly conserved by the Vlasov
equation, but it can be approximately conserved in some situations)
\footnote{This idea turns out to be similar to that introduced earlier
by Stiavelli \& Bertin (1987) with the conservation of an additional
quantity $Q=\int J^{\nu}|\epsilon|^{-3\nu/4}f d{\bf r}d{\bf v}$, see
Sec. \ref{sec_phen}.}. The extremization problem leads to a DF of the
form $f=f(\epsilon_{a})$ with $\epsilon=\epsilon+{J^{2}\over 2r_{a}}$
and $f'(\epsilon_{a})<0$ ($r_{a}$ is the anisotropy radius and
$J=|{\bf r}\times {\bf v}|$ the angular momentum). Furthermore, the
condition of maximum is probably related to a form of nonlinear
dynamical stability. These intuitive results have to be made more
rigorous. (ii) To use the turning point method, we have been led to
confine the system within a box. Physically, only the last point in
the series of equilibria that corresponds to a self-confined
structure unaffected by the box (the density $\rho(R)=0$ vanishes on
the boundary) is physically relevant (see, e.g., the complete
polytrope in Fig.
\ref{el4}). However, the whole curve is needed to determine whether
this state lies before or after the turning point of ``temperature''
$\beta$ or energy $E$. Certainly, the dynamical stability of
self-confined systems (with a compact support) could be settled by
other means, without plotting the whole series of equilibria for
box-confined configurations (see Rein 2005). However, in that case
we miss the interesting analogy with the thermodynamical problem
that we have found. Furthermore, there are situations where the box
is necessary. Technically, the box is needed when the system can
never be self-confined. This is the case for isothermal stellar
systems and polytropes with index $n\ge 5$. One justification of the
box in that case is to delimitate the region of space where the
distribution function of the above type holds \footnote{Instead of a
rigid box, we could consider a system in contact with an external
medium imposing a constant pressure $P$. In that case, the method of
linear series of equilibria still works in the so-called isobaric
ensemble (see Chavanis 2003a).}. For example, the box could mimic
the effect of a tidal radius in the case of globular clusters. On
the other hand, we have found that the dynamical and thermodynamical
methods developed in astrophysics can have applications for other
systems with long-range interactions like two-dimensional vortices
in hydrodynamics (Chavanis 2002c), bacterial populations in biology
(Chavanis et al. 2004) and the HMF model (Chavanis et al. 2005).
For these systems, there are situations where the introduction of a
material box is physically relevant. The general study of
the stability problems, even in idealized situations of box-confined
configurations, is not just academic but can have useful
applications in unexpected areas. Therefore, it is important to
consider the problem in full generality even if we sacrifice for
direct astrophysical applications. Finally, instead of introducing a
material box, we can use a confinement in energy space by using
truncated distribution functions like the Michie-King model. In that
case the method of linear series of equilibria still holds as shown
by Katz (1980) who analyzed different truncated models and
determined the limit of stability for each of them. Interestingly,
he found that the limits of stability are not too sensitive to the
form of confinement. Note that according to the criterion
(\ref{vh3}), the stability limits found by Katz (1980) can be
interpreted either as limits of thermodynamical stability or as
limits of nonlinear dynamical stability with respect to the
Vlasov-Poisson system.

In conclusion, the method of linear series of equilibria is very
powerful and allows us to determine the limits of stability for various
models. In particular, one goal of this paper was to show that this
method can be used for nonlinear dynamical stability problems, not
just for thermodynamical stability problems. This offers a
connection between different types of models: isothermal stellar
systems, self-gravitating fermions, polytropes, Michie-King models
(truncated isothermals) etc. Furthermore, they allow us to determine
their stability limits very easily from a purely graphical
construction: we just need to plot the series of equilibria $\beta(E)$
(this only demands that we solve the {\it first order} variational problem)
and then check whether the configuration of interest lies before or
after the turning point of temperature or energy in the series of
equilibria. In thermodynamics, the first turning point of temperature
implies a loss of canonical stability and the first turning point of
energy a loss of microcanonical stability. For the nonlinear dynamical
stability problem, the first turning point of ``temperature'' (in the
sense of our thermodynamical analogy) implies a loss of nonlinear
dynamical stability for a barotropic star with respect to the
Euler-Poisson system and the first turning point of energy a loss of
nonlinear dynamical stability for a spherical stellar system with
respect to the Vlasov-Poisson system.

\appendix

\section{The point of marginal stability}
\label{sec_marg}

Let us consider a stationary solution of the Euler-Poisson system
(\ref{ep1})-(\ref{ep3}) satisfying ${\bf u}={\bf 0}$ and the
condition of hydrostatic equilibrium (\ref{ep5}). The linearized
Euler-Poisson equations around this solution are
\begin{eqnarray}
\label{puls1} {\partial\delta \rho\over\partial t} +\nabla\cdot
(\rho\delta {\bf u})=0,
\end{eqnarray}
\begin{eqnarray}
\label{pusl2} \rho {\partial \delta {\bf u}\over\partial t}=
-\nabla(
p'(\rho)\delta\rho)-\rho\nabla\delta\Phi-\delta\rho\nabla\Phi,
\end{eqnarray}
\begin{eqnarray}
\label{pusl3} \Delta\delta\Phi=4\pi G\delta\rho.
\end{eqnarray}
Considering spherically symmetric systems and writing the evolution of
the perturbation as $\delta\rho\sim e^{\sigma t}$, we get
\begin{eqnarray}
\label{pusl4} \sigma\delta\rho+{1\over r^{2}}{d\over dr}
(r^{2}\rho\delta u)=0,
\end{eqnarray}
\begin{eqnarray}
\label{pusl5} \sigma\rho\delta u=-{d\over dr}(p'(\rho)\delta\rho)-
\rho {d\delta\Phi\over dr}-\delta\rho{d\Phi\over dr},
\end{eqnarray}
\begin{eqnarray}
\label{pusl6} {1\over r^{2}}{d\over dr}\biggl
(r^{2}{d\delta\Phi\over dr}\biggr )=4\pi G\delta\rho.
\end{eqnarray}
We write
\begin{eqnarray}
\label{pusl7} \delta\rho={1\over 4\pi r^{2}}{dq\over dr},
\end{eqnarray}
where $q(r)\equiv \delta M(r)=\int_{0}^{r}\delta\rho(r')4\pi
r'^{2}dr'$ denotes the mass perturbation within the sphere of radius
$r$. Using the Gauss theorem $d\delta\Phi/dr=Gq/r^{2}$, we see that the
Poisson equation (\ref{pusl6}) is automatically satisfied. On the other hand, the continuity equation (\ref{pusl4}) leads to
\begin{eqnarray}
\label{pusl8} \delta u =-{\sigma \over 4\pi \rho r^{2}}q,
\end{eqnarray}
where we have used $q(0)=\delta u(0)=0$ to eliminate the constant of
integration. Finally, substituting Eqs. (\ref{pusl7}) and
(\ref{pusl8}) in Eq. (\ref{pusl5}) and using the Gauss theorem and the
condition of hydrostatic equilibrium $dp/dr=-\rho d\Phi/dr$ we finally
obtain
\begin{equation}
{d\over dr}\biggl ({p'(\rho)\over 4\pi \rho r^{2}}{dq\over dr}\biggr
)+{Gq\over r^{2}}={\sigma^{2}\over 4\pi \rho  r^{2}} q.
\label{pusl9}
\end{equation}
This is the required equation of pulsations for a self-gravitating
barotropic gas. It is equivalent to the Eddington (1926) equation of
pulsations albeit written in a different form (see Chavanis \& Sire
2005). For isothermal and polytropic equations of state, we recover
the equations studied in Chavanis (2002ab,2003a). It can be shown that
the eigenvalues $\sigma^{2}$ are real. Therefore, the condition of
stability is $\sigma^{2}<0$ corresponding to an oscillatory behavior
of the perturbation. The condition of marginal stability is
$\sigma=0$.

Let us now consider the minimization of the energy functional
(\ref{ep4}). The second order variations can be written
\begin{equation}
\delta^{2}{\cal W}={1\over 2}\int_{0}^{R}{dq\over dr}\delta\Phi dr+\int_{0}^{R}{p'(\rho)\over 8\pi \rho r^{2}}\left ({dq\over dr}\right )^{2}dr.
\label{pusl10}
\end{equation}
Integrating by parts and using $q(0)=q(R)=0$, we obtain
\begin{eqnarray}
\delta^{2}{\cal W}=-{1\over 2}\int_{0}^{R}q {d\delta\Phi\over dr} dr-\int_{0}^{R}{d\over dr}\left ({p'(\rho)\over 8\pi \rho r^{2}}{dq\over dr}\right ) q dr.
\nonumber\\
\label{pusl11}
\end{eqnarray}
Then, using the Gauss theorem, we get
\begin{eqnarray}
\delta^{2}{\cal W}=-{1\over 2}\int_{0}^{R}\left\lbrack {d\over dr}\biggl ({p'(\rho)\over 4\pi \rho r^{2}}{dq\over dr}\biggr
)+{Gq\over r^{2}}\right\rbrack q dr.
\label{pusl12}
\end{eqnarray}
We are led therefore to consider the eigenvalue problem
\begin{eqnarray}
\left \lbrack {d\over dr}\biggl ({p'(\rho)\over 4\pi \rho r^{2}}{d\over dr}\biggr
)+{G\over r^{2}}\right\rbrack q_{\lambda}(r)=\lambda q_{\lambda}(r).
\label{pusl13}
\end{eqnarray}
If all the eigenvalues $\lambda$ are negative, then $\delta^{2}{\cal
W}>0$ and the configuration is a minimum of ${\cal W}$ at fixed
mass. This implies that it is nonlinearly dynamically stable. If at
least one eigenvalue $\lambda$ is positive, the configuration is a
saddle point of ${\cal W}$ and the stability criterion is not
satisfied. The marginal case is when the largest eignenvalue $\lambda$
is equal to zero. In that case, Eqs (\ref{pusl9}) and (\ref{pusl13})
coincide. This implies that the point in the series of equilibria
where the system becomes linearly unstable $(\sigma=0)$ coincides with
the point where it ceases to be a minimum of ${\cal W}$, i.e.
$\lambda=0$. Therefore, the onset of linear and nonlinear dynamical
stability coincides, as previously found in the particular cases of
isothermal and polytropic distributions (Chavanis 2002ab,2003a).

\section{Passage from $F[f]$ to $F[\rho]$}
\label{sec_pass}

The optimal distribution function $f_{*}$ which minimizes $F[f]$ at
fixed density profile $\rho({\bf r})$ is determined by
$C'(f_{*})=-\beta({v^{2}/2}+\lambda({\bf r}))$. Since $C$ is
convex, this relation can be inversed to give
\begin{equation}
\label{pass1}f_{*}({\bf r},{\bf v})=F\left\lbrack \beta \left ({v^{2}\over 2}+\lambda({\bf r})\right )\right\rbrack,
\end{equation}
where $F(x)=(C')^{-1}(-x)$. The density $\rho=\int f_{*}d{\bf v}$ and the pressure $p={1\over d}\int f_{*}v^{2}d{\bf v}$ (in $d$ dimensions) can then be expressed as
\begin{equation}
\label{pass2}\rho={1\over \beta^{d/2}}g(\beta\lambda), \qquad p={1\over \beta^{d+2\over 2}}h(\beta\lambda),
\end{equation}
with
\begin{equation}
\label{pass3} g(x)=2^{d-2\over 2}S_{d} \int_{0}^{+\infty} F(x+t)\ t^{d-2\over 2} dt,
\end{equation}
\begin{equation}
\label{pass4} h(x)={1\over d}2^{d\over 2}S_{d} \int_{0}^{+\infty} F(x+t)\ t^{d\over 2} dt,
\end{equation}
where $S_{d}$ is the surface of a unit sphere in $d$
dimensions. Eliminating $\lambda$ between the foregoing expressions,
we find that $p=p(\rho)$ where the barotropic equation of state is
entirely specified by $C(f)$. We can now express $F[f]=E[f]-TS[f]$ as
a functional of $\rho$ by writing $F[\rho]=F[f_{*}]$. The energy 
is simply given by
\begin{equation}
\label{pass5}
E={d\over 2}\int p\ d{\bf r}+{1\over 2}\int\rho\Phi d{\bf
  r}.
\end{equation}
On the other hand, the Casimir (\ref{vh2}) can be written
\begin{eqnarray}
{S}=-{2^{d-2\over 2}S_{d}\over\beta^{d/2}}\int d{\bf
  r}\int_{0}^{+\infty}C\lbrack F(t+\beta\lambda)\rbrack \ t^{d-2\over 2}dt.
\label{pass6}
\end{eqnarray}
Integrating by parts  and using $C'\lbrack F(x)\rbrack=-x$, we find that
\begin{eqnarray}
{S}=-{2^{d/2}S_{d}\over d\beta^{d/2}}\int d{\bf r}\int_{0}^{+\infty}
F'(t+\beta\lambda)(t+\beta\lambda)t^{d/2}dt.
\label{pass7}
\end{eqnarray}
Integrating by parts one more time and using Eqs. (\ref{pass2}),
(\ref{pass3}) and (\ref{pass4}), we finally obtain
\begin{eqnarray}
{S}={d+2\over 2}\beta\int p d{\bf r}+\beta\int \lambda\rho d{\bf r}.
\label{pass8}
\end{eqnarray}
Collecting all the previous results, we get
\begin{eqnarray}
{F}[\rho]=-\int\rho \biggl (\lambda+{p\over\rho}\biggr )d{\bf
r} +{1\over 2}\int\rho\Phi d{\bf r}. \label{pass9}
\end{eqnarray}
Finally, using the relation $h'(x)=-g(x)$ obtained from
Eqs. (\ref{pass3}) and (\ref{pass4}) by a simple integration by parts, it
is easy to check that Eq. (\ref{pass2}) implies
\begin{eqnarray}
\lambda'(\rho)=-{p'(\rho)\over\rho},
\label{pass10}
\end{eqnarray}
so that
\begin{eqnarray}
\lambda+{p\over\rho}=-\int^{\rho}{p(\rho')\over\rho'^{2}}d\rho'.
\label{pass11}
\end{eqnarray}
Thus,
\begin{eqnarray}
{F}[\rho]=\int \rho \int_{0}^{\rho}{p(\rho')\over\rho'^{2}}d\rho'
d{\bf r}+{1\over
  2}\int\rho\Phi d{\bf r}.
\label{pass12}
\end{eqnarray}

\end{document}